\documentclass[preprint1]{aastex}

\usepackage{ulem} 

\usepackage{longtable}
\usepackage{rotating}

\usepackage{ulem}

\shorttitle{CTE in ACS}
\shortauthors{Anderson \& Bedin}

\begin{document}


\title{An Empirical Pixel-Based Correction for Imperfect CTE.\ \ I.\ \ 
       HST's Advanced Camera for Surveys\footnote{
               Based on observations with the NASA/ESA 
               {\it Hubble Space Telescope}, obtained at the 
               Space Telescope Science Institute, which is operated by 
               AURA, Inc., under NASA contract NAS 5-26555.}} 

\author{Jay Anderson {\it and\ } Luigi R. Bedin}

\affil{Space Telescope Science Institute,
       Baltimore, MD 21218, USA; jayander,bedin@stsci.edu}

\received{{\it 2010 March 23}}
\accepted{{\it 2010 July  21}}

\begin{abstract}

We use an empirical approach to characterize the effect of charge-transfer 
efficiency (CTE) losses in images taken with the Wide-Field Channel 
of the Advanced Camera for Surveys.  The study is based on profiles 
of warm pixels in 168 dark exposures taken between September and 
October 2009.  The dark exposures allow us to explore charge traps 
that affect electrons when the background is extremely low.  We 
develop a model for the readout process that reproduces the observed 
trails out to 70 pixels.  We then invert the model to convert the 
observed pixel values in an image into an estimate of the original 
pixel values.  We find that when we apply the image-restoration 
process to science images with a variety of stars on a variety of 
background levels, it restores flux, position, and shape.  This means 
that the observed trails contain essentially all of the flux lost to 
inefficient CTE.  The Space Telescope Science Institute is currently 
evaluating this algorithm with the aim of optimizing it and eventually 
providing enhanced data products.  The empirical procedure presented 
here should also work for other epochs (eg., pre-SM4), though the 
parameters may have to be recomputed for the time when ACS was operated 
at a higher temperature than the current $-81^{\circ}$ C.  Finally, 
this empirical approach may also hold promise for other instruments, 
such as WFPC2, STIS, the ACS's HRC, and even WFC3/UVIS. 

\end{abstract}



\section{INTRODUCTION}
\label{s.INTRO}
\begin{figure}
\plotone{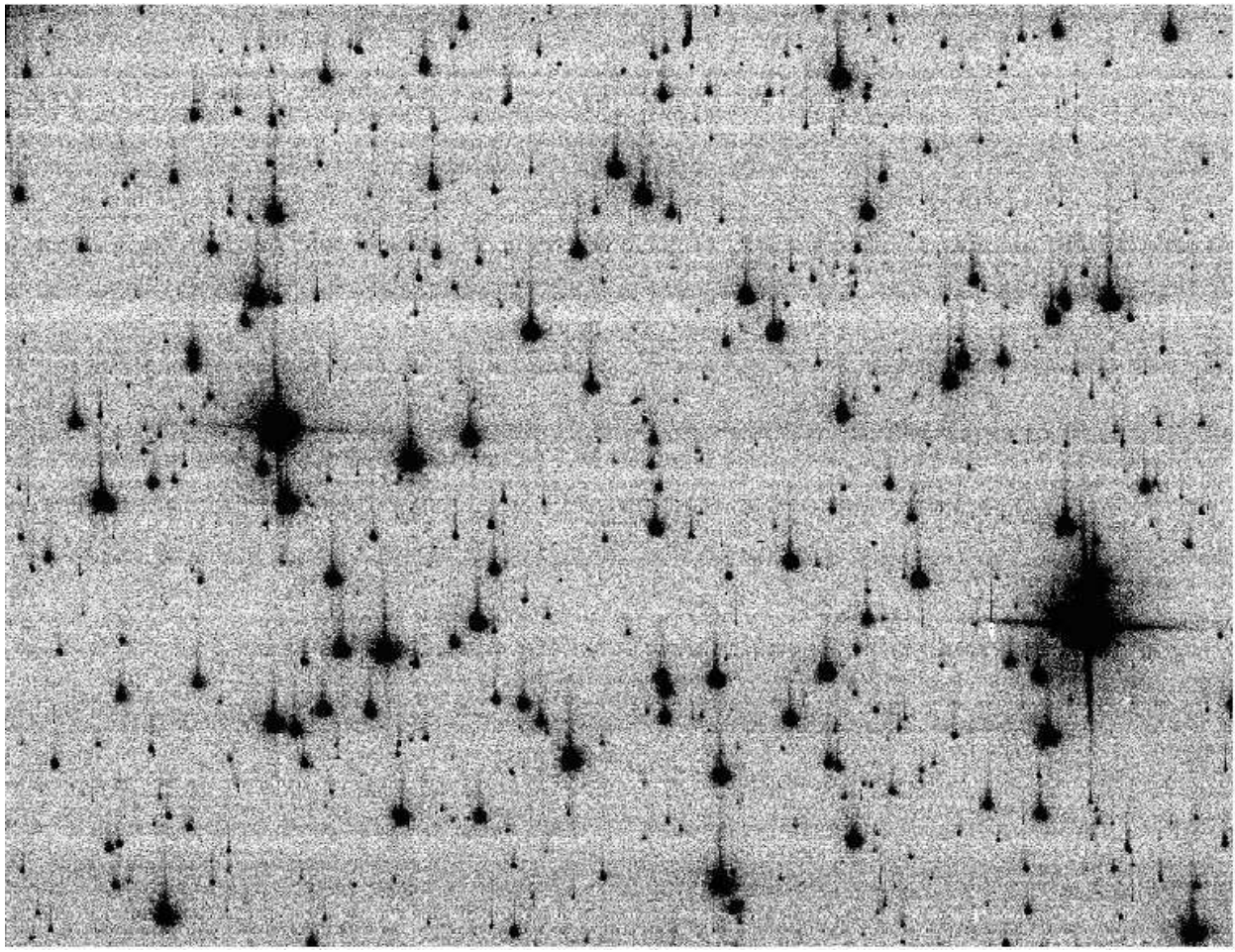}
\caption{A 850$\times$700-pixel region centered on (1650,1725) of
         the bottom chip in image {\tt ja9bw2ykq\_flt}, a 30 s
         exposure of the 47 Tuc calibration field.  The background
         is $\sim$ 3 $e^{-}$.  The vertical trails extending 
         upward from the stars are indicative of imperfect CTE; the 
         horizontal streaks are an artifact of the post-repair readout 
         electronics and are currently under study (Grogin et al.\ 2010). 
         \label{fig01}}
\end{figure}

It is well known that when energetic particles impact CCD detectors, 
they can displace silicon atoms and create vacancies (defects) in the 
silicon lattice.  During the read-out process, these defects can 
temporarily trap electrons, causing some of a pixel's charge to arrive 
late at the read-out register, and thus to be associated with a different 
pixel.  These transfer imperfections are generally referred to as CTE 
(charge-transfer efficiency) losses or CTI (charge-transfer inefficiency), 
since some flux that was initially in one pixel gets delayed during 
the transfer process and shows up in a pixel that is read out later, 
sometimes much later.  Detectors in the harsh radiation environment 
of space suffer this degradation much faster than detectors on the 
ground, making it a particularly serious problem for the aging instruments 
on board the Hubble Space Telescope (HST).  The ubiquitous CTE-related 
trails stand out clearly in post-repair images taken with the Advanced 
Camera for Survey's Wide Field Channel (ACS's WFC).   Figure~\ref{fig01} 
shows the upward-streaking trails in a recent short exposure.

Imperfect CTE can have 
two main impacts on science.  First, it shifts charge from the 
core of a source to a trail that extends well outside of a normal 
photometric aperture, thus reducing the brightness measured in 
aperture photometry and PSF-fitting.  This has a significant impact 
on photometry, both for point sources and for extended sources.  Second, 
imperfect CTE blurs out the profiles of objects, shifting their centroids 
and increasing their FWHMs.  This has a major impact on astrometric 
projects and on studies that involve galaxy morphology, such as weak 
lensing.

Because of the detrimental impact of imperfect CTE on science, 
much effort has been expended in order to understand what causes it and 
how to characterize it, both in the laboratory and by studying images 
taken with detectors that have suffered actual space-radiation damage.  
Janesick (2001, Chapter 8) summarizes the recent laboratory-based 
research into what damage various kinds of radiation can inflict on 
CCDs.  Although it is hard to simulate in the laboratory the exact 
radiation environment of space, it is possible to construct general 
models for how defects are generated and how they impact the transfer 
of charge from the pixels to the readout register.  Laboratory testing 
has demonstrated that defects tend to capture electrons very quickly, 
but release them much more slowly, all with time constants that depend 
on temperature (see Hardy, Murowinski, \& Deen 1998).  The testing has 
also shown that charge packets with smaller numbers of electrons are 
exposed to a larger number of traps (per electron).  These 
characteristics are generally consistent with our observational experience
with HST, but unfortunately the laboratory tests are not yet sufficient 
to provide reliable quantitative predictions.  This is partly because 
the laboratory tests cannot fully simulate realistic space environments, 
and partly because the laboratory tests generally rely on a limited number 
of photon sources (such as X-rays from $^{55}$Fe) that generate fixed-size 
charge packets, and thus they are unable to probe the full parameter 
space of how charge packets of different sizes experience deferred 
charge.  Until laboratory results are able to predict and describe 
the full spectrum of on-orbit CTE losses more quantitatively, we must 
develop empirical approaches to characterize the specific detectors 
currently operating in space.  These approaches should be guided by 
the many insights gleaned in the laboratory.

Several attempts have been made over the years to empirically characterize 
the impact of imperfect CTE on WFC observations.  A series of Instrument 
Science Reports (ISRs) (see Riess and Mack 2004, Chiaberge et al.\ 2009, 
and references therein) have developed empirical approaches to correcting 
aperture photometry.  These corrections depend on the size of the 
aperture, the brightness of the source, the intensity of the background, 
and the time since exposure to the harsh environment.  The losses are 
lower when the background is higher, which is consistent with the 
laboratory results that show that some of the charge traps can be filled 
by the background and are thus not seen by sources.  These reports 
also demonstrate very clearly that the amplitude of CTE losses has been
increasing linearly with time, also in accord with lab results 
(see Fig.~8 of Waczynski et al 2001).  Dolphin has made similar CTE 
characterizations for the WFPC2 detectors (see Dolphin 2000 and 2009).

The empirical photometric-correction trends with magnitude are 
sufficient for many applications, but they have their limitations.
In practice the corrections are constructed only for a limited set 
of photometric apertures, which does not allow easy correction for stars 
in crowded fields or for faint stars near the background, where even a 
3-pixel-radius aperture can lead to unacceptably large systematic or 
random errors (see Anderson et al.\ 2008).  Furthermore, the corrections 
constitute average corrections and do not take into account the particular 
shadowing circumstances of each star (i.e., the fact that some stars may
sit on higher backgrounds than others, or may have well-placed neighbors 
that may sheild them from CTE losses).  In addition, the photometric trends 
tell us nothing about how astrometry or an object's shape may have been 
impacted by imperfect CTE.  For all these reasons, over the years many 
attempts have been made to come up with a pixel-based CTE-correction:   
a way to reconstruct the original image based on the observed image and 
a model for CTE. 

Riess (2000) studied empirical galaxy profiles in WFPC2 images and 
constructed a model of the readout that reproduced the phenomenology he 
observed in images, but the algorithm was too slow to be used in practice 
(Riess, personal communication).  For STIS, Bristow \& Alexov (2002) and 
Bristow (2003a, 2003b) developed a theoretical model for how charge is 
read out from the detector, based on lab experiments and a theoretical 
charge-trapping and -emission model from Philbrick (2001). They used 
experimental data, such as trap densities, release time constants, and 
the design specs for a ``notch'' channel (which was intended to insulate 
the smallest charge packets from CTE losses) to construct a computer 
program that simulated the STIS readout.  They use this routine to generate 
a realistic simulation model of lab data, then used the model to generate 
a pixel-by-pixel correction for science images that did a good job 
qualitatively removing the CTE trails.  Nevertheless, despite this 
\ae sthetic success, the algorithm was not able to restore 
all the flux to the rightful pixels and was too computationally slow 
to be practical, so the standard CTE-mitigation procedure for STIS 
to this day still involves measuring quantities on the uncorrected 
pixels and applying parametrized corrections based on background, 
source flux, and aperture size.  This is similar to what is done for 
ACS and WFPC2.  We note that, unlike ACS, STIS can be read out using the 
amplifiers at any of its four corners, which makes its CTE much 
easier to measure and evaluate (see Goudfrooij et al.\ 2006).
 
More recently, Massey et al.\ (2010) have constructed a pixel-based CTE 
correction for ACS's WFC for their reductions of the COSMOS fields.  They 
use power laws and exponentials to model the amplitude and profiles of the
CTE-related trails behind warm pixels (hereafter, WPs).  They find that 
their pixel-correcting procedure is able to remove the visible trails and 
reconstruct the images of their targets of interest (resolved galaxies), 
with the result that the galaxies near the top of the chips, which
are farthest from the readout amplifiers and naturally suffer 
more CTE losses, have shapes after correction that are similar to those 
at the bottom of the chips, which suffer less CTE losses.  The 
Massey et al.\ correction was tailor-made for the particular 
characteristics of the COSMOS field:  a flat background of about 
50 $e^{-}$ and relatively faint sources.  They make no claims as 
to how well their algorithm would perform on the full range of 
backgrounds and source brightnesses of fields studied with the WFC.  
It is also not clear whether their corrections yield accurate 
photometry and astrometry, since they did not have access to true 
fluxes and positions against which they could compare their results.
Nevertheless, the clear success of their algorithm at removing the trails 
and restoring flux to galaxy profiles is an extremely encouraging 
indication that a pixel-based correction for CTE is possible.

To that end, we decided to carry their study to the next level.  Our 
aim is to characterize the impact of imperfect CTE over the full range 
of background intensity and source flux.  
Our plan will be as follows.

Although Massey et al.\ (2010) were able to successfully model the 
trails in terms of exponentials with experimentally motivated drop-offs, 
we will deal with the trails in a purely empirical, non-parametric
way.  We will constrain various aspects of our model by studying how CTE 
impacts WPs and cosmic rays (CRs).  The WPs in dark frames serve as 
delta functions that allow us to calibrate the size and extent of CTE 
trails for pixels of various flux in images that have essentially no 
background\footnote{
      Note we will refer here to all pixels with a discernible 
      dark-current excess as ``warm'', irrespective of whether the 
      ACS Instrument Handbook would formally characterize them as 
      ``hot''($> 0.08 ~ e^{-}$/pix/s), 
      ``warm'' (0.04 to 0.08 $e^{-}$/pix/s), 
      or below these thresholds.}.
The CRs allow us to examine the effects of ``shadowing'' on CTE.  
It is well known that point sources suffer lower CTE-related losses
when the background is higher, but it is unclear whether the
whole-chip background or the very local background is more relevant
to this trend.  Laboratory experiments (see Hardy et al.\ 1998)
suggest that this is related to the trap-capture time constants 
for the detector, which should be very fast and almost instantaneous
relative to the parallel pixel clock-time.  We will use the trails
from CRs to tell us whether the trap-time truly is instantaneous.

Once the shadowing properties have been characterized, we will turn 
our attention to the warm pixels and construct an empirical model
of the WFC pixel-readout process that reproduces the observed tails 
behind the delta-function warm pixels.  With such a model in hand, it 
will be straightforward to construct an iterative forward-modeling 
procedure to restore the observed flux to its rightful pixel.  Finally, 
the examination of stars will provide the ultimate test.  By comparing 
corrected fluxes and positions for stars in short and deep exposures, 
we will demonstrate that this correction procedure works for all 
backgrounds and fluxes.

This paper is organized as follows.  In Section~\ref{s.DARK}, we 
first analyze a set of 168 calibration images that were taken between 
September and October 2009 to measure the dark current.  We produce a 
stacked dark image and maps of where the persistent warm pixels are 
located.  The stacks provide a high S/N profile of each charge trail, 
and we combine the many trails to examine trends in trail-profile shape 
and intensity as a function of WP intensity.  In Section~\ref{s.MODEL}, 
we develop an empirical model of the readout process, putting together 
everything we have learned thus far, from laboratory experiments,
from previous ISRs, papers, and our own qualitative analysis.  The 
parameters of this model are optimized empirically by fitting the 
warm-pixel trails in the dark exposures.  We test the efficacy of 
the algorithm for warm pixels in images with different backgrounds.  
In Section~\ref{s.STARS}, we present the stellar tests, demonstrating 
that the algorithm properly returns flux to the rightful pixels so that 
astrometry, photometry, and even shape are restored.  Section~\ref{s.IMPR}
discusses some improvements that could be made to make the algorithm 
better.  Finally, Section~\ref{s.SUMMARY} summarizes these results and 
considers the next logical steps to take.


\section{DARK-CURRENT OBSERVATIONS}
\label{s.DARK}

Most of HSTs orbits experience some occultation by the earth.  The 
observatory regularly takes advantage of this ``down'' time to collect
a rich assortment of bias frames, dark frames, and flats.  As a result, 
the archive contains an enormous number of such calibration data sets.  
Over the 55 days between 4 September 2009 and 28 October 2009, there 
were 168 dark exposures taken.  These dark exposures were between 
1000 s and 1090 s in length, with an average exposure time of 1061.2 s.
The darks were generally paired with bias exposures, making it easy 
to select a set of darks and biases for the same period of time.

\subsection{The Dark-Current Dataset}
\label{ss.DARK_DATA}

\begin{figure}
\plotone{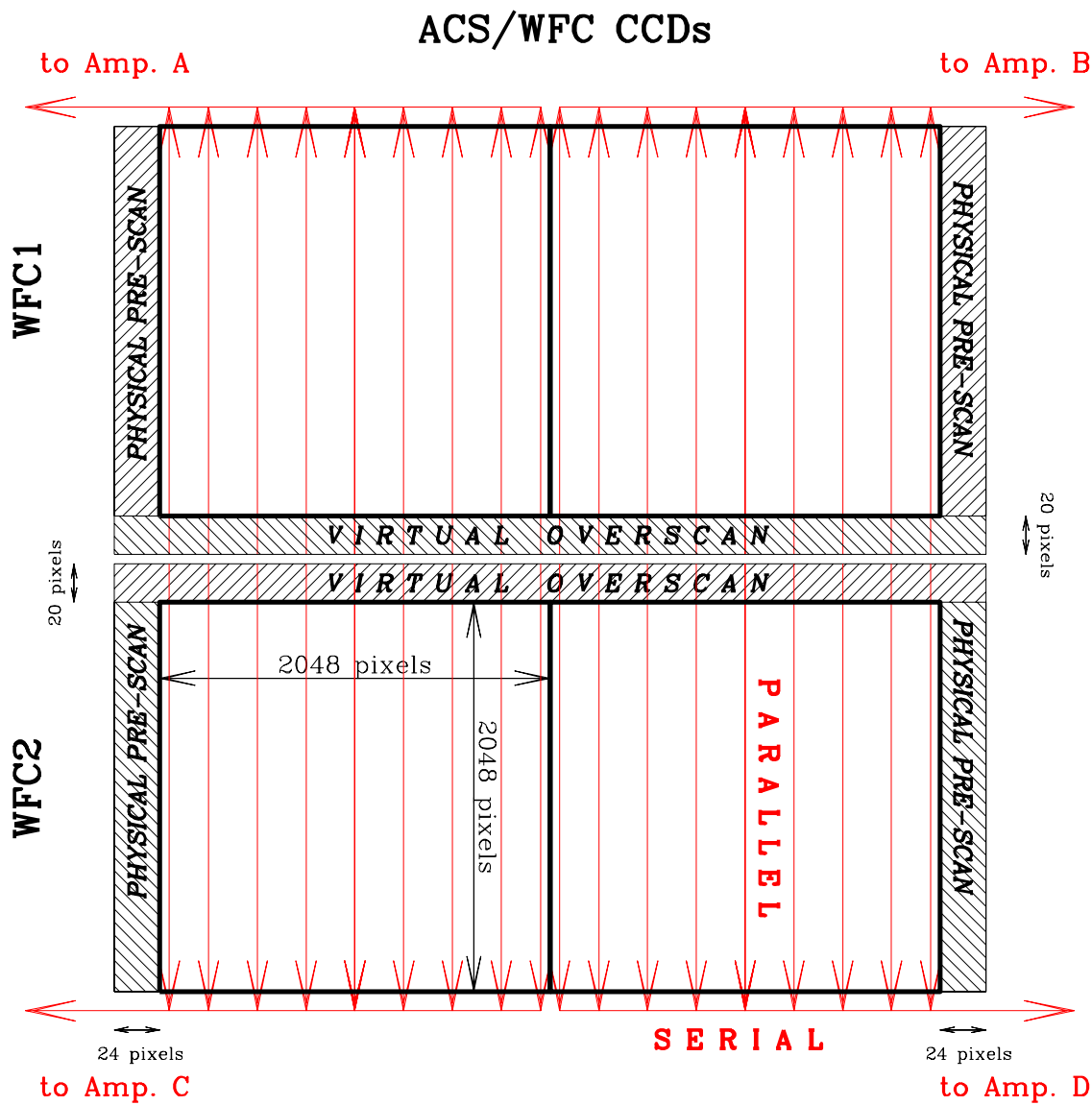}
\caption{The set-up of the physical pre-scan and virtual overscan regions 
         in the 4144 $\times$ 4136 {\tt \_raw} images, relative to the 
         readout directions.  The charge is first shifted along the
         parallel direction into the serial register, then this charge
         is shifted horizontally into the readout amplifiers, labeled 
         A, B, C and D.  The {\tt \_flt} images include only the 
         4096 $\times$ 4096 white region.  
         \label{fig02}}
\end{figure}

We will begin our investigation using the {\tt \_raw} dark-current
observations.  These are 4144 $\times$ 4136 two-byte unsigned-integer 
images that contain the raw number of counts recorded by the A/D 
converter for each pixel.  Figure~\ref{fig02} provides a useful 
schematic for these extended images.  The physical pre-scan on the 
right and left corresponds to 24 actual pixels that are not exposed 
to the sky but are still read out like normal pixels.  The pre-scan pixels 
are closest to the read-out amplifier for each quadrant.  The 20 pixels 
of virtual overscan (in the middle of the figure, farthest from the 
serial registers at the top and bottom) are not real pixels, but rather 
come from 20 extra parallel clockings of the rows.  The charge 
packets corresponding to these pixels start with no charge, but can gain 
charge on account of electrons released from CTE-traps or CRs that 
impact during readout.  From Mutchler \& Sirianni (2005, p3), we 
find that the serial shift time is 22 $\mu$s, and the parallel shift 
time is 3212 $\mu$s, so that the entire array is read out in 
(20+2048) $\times$ (3212 + 22$\times$(2048+24)) $\mu$s, 
or about 100 seconds.


\begin{figure}
\plotone{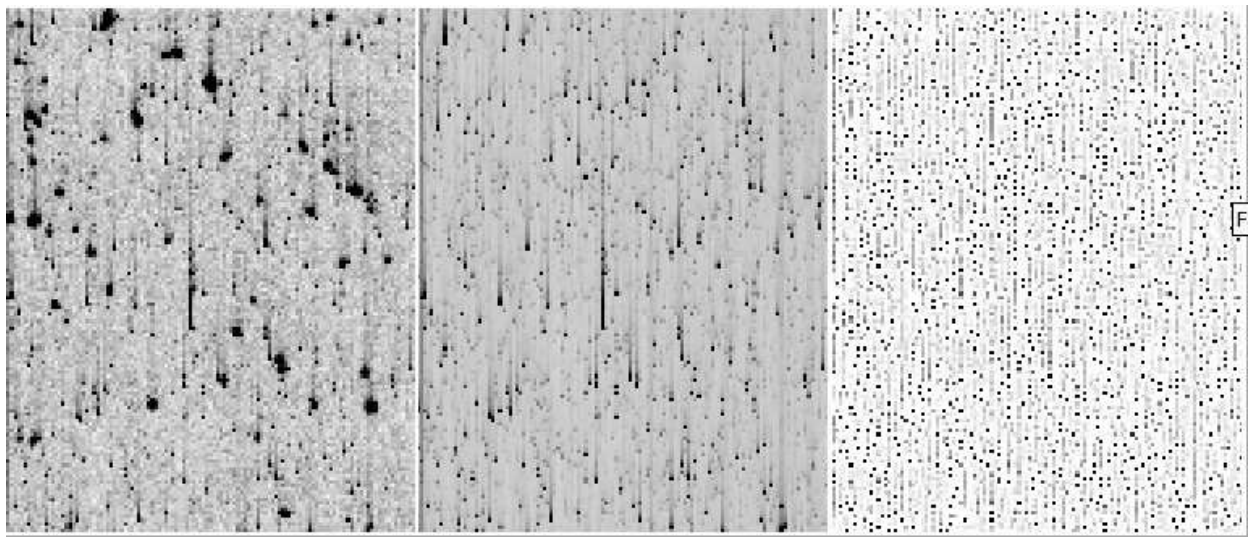}
\caption{We show a 50 $\times$ 75 pixel region near the top of WFC2
         (the bottom chip).  The left panel shows a single dark exposure 
         ({\tt jbanaan2q}).  The middle panel shows the average for 
         all the 168 darks considered.  The right panel shows the 
         corresponding peak map.  The darkest points correspond to 
         pixels found to be local maxima in more than 150/168 
         exposures.  The trails show up faintly in the peak map, as 
         they are slightly more likely that other pixels in their row 
         to be local maxima.
         \label{fig03}}
\end{figure}

In addition to trapping electrons during readout, defects 
      caused by radiation damage also allow electrons to ``hop''
      from the silicon valence band to the conduction band, thus 
      increasing the dark current (Sirianni et al 2007).  The 
      typical post-SM4 dark current in the WFC is ~0.006 e$^{-}s^{-1}$
      per pixel, but after many years of exposure to radiation, every 
      pixel has a different radiation history and a different 
      dark-pixel intensity.  About 10\% of the pixels have more than 
      0.015 e$^{-}s^{-1}$, and 2\% have more than 0.05 e$^{-}s^{-1}$.
The plentiful dark exposures 
contain a wealth of data about 
the detector defects and about the cosmic-ray intensity and frequency.
The left panel of Figure~\ref{fig03} shows a single dark exposure 
({\tt jbanaan2q}, the first one in our list).  In general, WPs are 
the single-pixel events and cosmic rays are the multiple-pixel 
events, with the typical CR affecting about 8 pixels.  The field shown 
covers a region near the top of WFC2 (the bottom chip), and is 
therefore about 2000 pixels from the bottom serial register.  As such, 
this region suffers maximally from CTE losses and the WP trails are quite 
obvious.


\subsection{Image analysis}
\label{ss.DARK_ANAL}

We will use these 168 images in two ways.  First we will distill 
them into two useful composite images that will help us study the 
average behavior of the WPs.  Then later we will study them individually
to understand what the transient component (namely, CRs) can
tell us.

The first useful composite image we generated is a simple stack that shows 
the average dark frame.  We construct this by normalizing the 
168 {\tt \_raw} images to the average exposure time of 1060 s, then 
for each of the 4144 $\times$ 4136 pixels, we list the 168 values 
and compute a robust average value by iteratively clipping the 
values beyond 4-$\sigma$.  The result of this is the image shown 
in the middle panel of Figure~\ref{fig03}.  The CRs are gone, and 
the trails from the WPs are clear.

The second useful image is what we call a ``peak map''.  To construct 
this map, we go through each individual dark exposure and identify all 
the local maxima (any pixel greater than its 8 surrounding neighbors).
The peak map records how many of the 168 images have a peak in each 
of the 4144 $\times$ 4136 pixels.  This map is shown in the right 
panel of Figure~\ref{fig03} and identifies the locations of the 
persistent warm pixels.  There are $3.3 \times 10^5 $ WPs 
found in 150 or more out of 168 exposures, $5.3 \times 10^5$ found 
in more than 125, and $7.2 \times 10^5$ found in more than 100.  
The WPs selected in this way typically have intensities of 
15 DN$_2$ or more\footnote{
      We note here that the {\tt \_raw} frames were all taken with 
      {\tt gain=2} so that the images could probe the entire dynamic 
      range of $\sim$85,000 e$^{-}$.  As such, each count corresponds 
      to 2 electrons.  To emphasize this, we refer to the count 
      units as DN$_2$, when appropriate.  
      In general, the analysis in this section and the beginning
      of Section 3 will use the {\tt \_raw} images and  will 
      involve DNs, but beginning in Section 4, we will study 
      the {\tt \_flt} images, where electrons are the natural unit.
      }.
Because of crowding and read-noise, the WPs much fainter than 
this do not generate peaks in more than 100/168 exposures, 
making them hard to study this way.

We made an initial bias correction for each frame using the physical 
pre-scan pixels.  Unfortunately, this does not remove all the bias, 
so we processed the contemporaneous bias {\tt \_raw} frames as above 
for the darks, and generated an average image of the bias.  This 
average image contains a significant gradient, which increases
in $x$ and $y$ away from the readout registers.  According to
Golimowski et al.\ (2010), these $x$ and $y$ gradients are due to
a ``bias drift'' in the readout voltage, and are not indicative of 
actual electrons being transferred by the detector.  Therefore, we simply 
subtracted this average bias from the dark frame before continuing 
with the analysis.

\subsubsection{Quantitative trails}
\label{ss.DARK_TAIL}

We used the peak map to identify the consistently warm pixels (those 
that stood out in at least 100 out of 168 exposures).  In order to
maximize the CTE signature, we examined the pixels that were at least
1500 pixels from the serial registers:  $j>1500$ in the WFC2 (the 
bottom CCD) and $j<500$ in the WFC1 (the top CCD).

\begin{figure}
\plotone{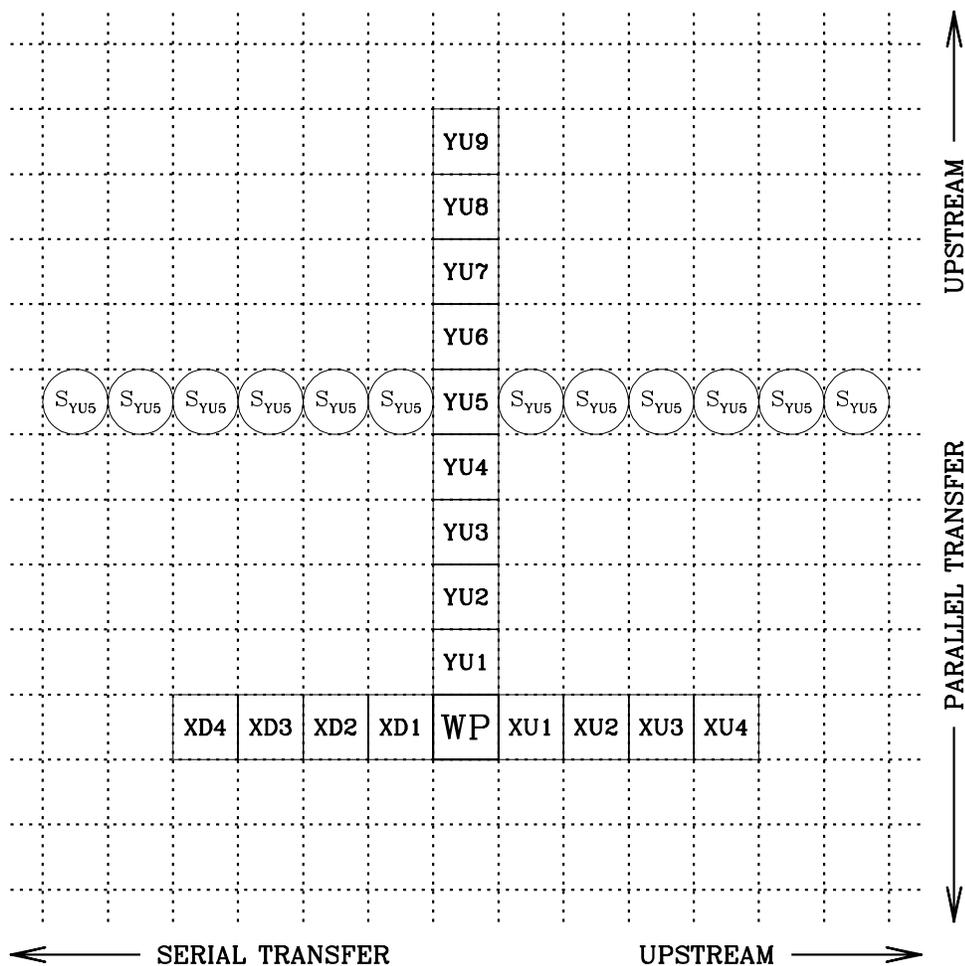}
\caption{This schematic shows the pixels used to construct the
         CTE-release trails.  The WP label indicates the location of the 
         warm pixel.  The parallel and serial transfer directions
         are indicated on the right and bottom, respectively.  
         The YUn pixels are the upstream pixels that make up the 
         Y-CTE trail.  The circled pixels are used to compute
         a sky-background correction for the fifth upstream pixel, YU5.  The
         upstream and downstream pixels in $x$ are also indicated. 
         \label{fig04}}
\end{figure}

The next step is to measure the CTE trails.  The schematic in 
Figure~\ref{fig04} illustrates which pixels we used to extract the 
CTE-release profiles upstream of each warm pixel.  The profiles have 
been zeropointed by subtracting a ``sky'' value for each pixel in the 
trail by taking a sigma-clipped average of the ten surrounding pixels 
in its row (this conveniently mitigates the impact of the horizontal 
streaks caused by $1/f$ noise in the new signal-processing electronics, 
see Grogin et al.\ 2010). 


\begin{figure}
\plotone{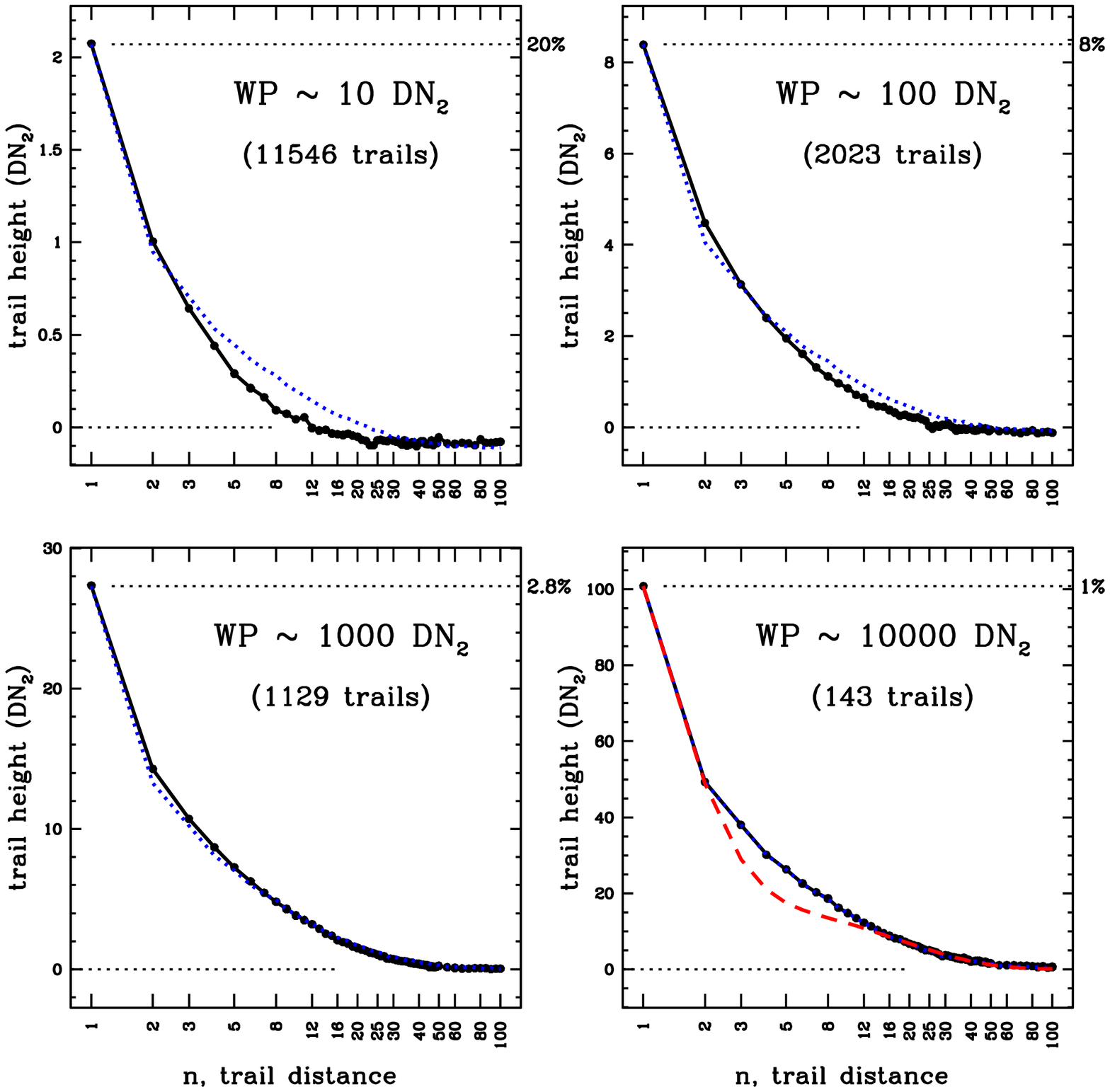}
\caption{The average extracted profiles for trails behind warm pixels 
         with intensities in the range of 10$\pm$1, 100$\pm10$, 
         1,000$\pm200$, and 10,000$\pm$2500 DN$_2$ that were at least 
         1500 pixels from the readout registers.  The fraction of flux 
         in the first upstream pixel (YU1) goes from 20\% to 1\%.  
         The dotted blue curve shows the WP=10,000 profile, scaled 
         to best match the other panels.  The red dashed curve in the 
         lower-right panel shows a dual-exponential fit constrained
         to match the inner and outer parts of the profile.
         \label{fig05}}
\end{figure}

Figure~\ref{fig05} shows the trails quantitatively.  The CTE correction
we implement will be based directly on profiles like these and will 
be tested by its ability to remove them from images.  The trails have 
been grouped in terms of the intensity of the warm pixel (labeled at 
the upper right in each panel).  The pixel value for the trail is listed 
on the left.  The percentage of WP flux in the first pixel in the trail 
is listed on the right.  This percentage goes from 20\% for low-intensity 
WPs to 1\% for WPs that reach a good fraction of full-well.  The fraction 
in the second pixel is down by about a half from that in the first pixel.

In addition to the fraction of flux in the trail changing dramatically 
with WP intensity, the shape also appears to change with WP intensity.  
The blue dotted line corresponds to the curve for 10,000 DN$_2$, and it
has been scaled to match the YU1 flux in the different profiles.  
While the trails do have the same general monotonic shape, the exact 
shape can differ by more than 10\%.  When we integrate this difference 
down the trail (note that the horizontal scale in Figure~\ref{fig05} is 
logarithmic), we find that this could account for several tens of 
percent in trail flux.

The red curve in the bottom right panel fits the 10,000 DN$_2$ profile 
with dual exponentials, the inner exponential constrained to fit the 
inner two points, and the outer exponential constrained to fit the 
outer half of the points.  It is clear that this does a very poor job 
representing the trail in the middle region.  Modeling the trails with 
simple dual-exponentials could clearly introduce errors of several tens of 
percent in the total trail flux.  

Finally, we note that the zeropoint for the faint-WP curves dips 
systematically below zero beyond where the trail can be measured.  This 
is simply an artifact of the difficulty in defining a sky value to better
than 0.2 DN$_2$ in the images.  While the images are putatively empty, the 
interacting trails from all the warm pixels leave a vertically coherent 
background that does not correspond to a Gaussian distribution with a 
mean of zero.  This makes it difficult to accurately measure flux far
down the faint trails. 


\section{OUR EMPIRICAL MODEL}
\label{s.MODEL}

The previous pixel-based CTE analyses discussed in the Introduction 
have assumed a single trail shape for all pixel fluxes.  
Figure~\ref{fig05} shows that this does not appear to be the case:  
reality may be more complicated.  Previous models have also treated 
the profiles as exponentials --- or as a sum of two exponentials, 
since such a model is well motivated by theory and lab experiments
(Massey et al 2010).  The red curve in Figure~\ref{fig05} shows that 
a dual-exponential does not satisfactorily follow all the signal in 
the trails.  Some recent lab experiments (see p820 Janesick 2010 or 
Table 6 in Sirianni et al 2007) have shown evidence for more than
two trap species, some of which have relatively long release times.  
It is possible that the observed trails might be well represented 
by three exponentials (with different scalings), but for simplicity 
here we will model the profile empirically by tabulating what fraction 
of charge is retained and released as a function of the number of 
pixel-transfers.  While this approach may involve more parameters than 
are strictly necessary, such a model is still considerably over-constrained 
by the plentiful data.  Irrespective of how they are fit, the 
curves in Figure~\ref{fig05} clearly demonstrate that CTE losses 
follows regular and predictable trends.

\subsection{Basic Constraints}
\label{ss.MODEL_BASIC}

Our modeling strategy will also take into account several basic 
facts about CTE that we gather from the extensive number of 
reports and papers that have been written about it:  (1) CTE 
losses are lower when the background is higher; (2) CTE losses 
are proportionally higher for fainter sources than for brighter 
sources; (3) CTE losses are directly proportional to the distance 
from the read-out register; and (4) CTE losses are directly 
proportional to the amount of time the detector has spent in 
the space environment.

While our model is purely empirical and does not depend on any 
quantitative results of lab experiments it does help to consider 
their qualitative results to motivate the model.  It is clear 
that CTE losses are related to the presence of defects or ``traps'' 
in the silicon lattice of the CCD.  These traps are generated 
when energetic photons, electrons, neutrons and ions impact
the detector, thus naturally explaining the linear increase in 
CTE losses over time in the high-impact environment of space.  
When the electrons from a pixel in the array are shifted through 
a downstream pixel, a trap in the downstream pixel may capture an 
electron from the packet.  This electron will be released some 
time later, after the packet of electrons with which it began its 
journey has already been transferred down several pixels.  The 
trail profiles in Figure~\ref{fig05} give us a rough 
sense of how long (in terms of pixel-shift times) the typical trap 
holds the typical electron.  

In general, the amount of charge in the trail tells us how many traps 
were encountered as the electron packet was shifted from the initial 
pixel to the readout register.  Of course, the packet may have 
encountered many more faster-release traps that may give up the electrons 
before the charge is shifted.  We can ignore these fast traps here:  
they will not affect the image, and furthermore we cannot hope to measure 
them by studying the image.  On the other hand, traps that retain electrons 
longer than the chip-readout time will not produce measureable electrons 
in the trails, though they will still reduce the flux of the stars.  The 
presence of these traps can only be inferred by comparing photometry 
against some absolute standard.  Biretta \& Mutchler (1997) show that 
the WFPC2 detector appears to have some extremely long traps, as 
evidenced by charge that persists into the next exposure.  While no 
such extreme-late-release charge has been observed in the WFC, we 
will certainly examine our photometry for losses not accounted
for by the flux in the observed trails.

\begin{figure}
\plotone{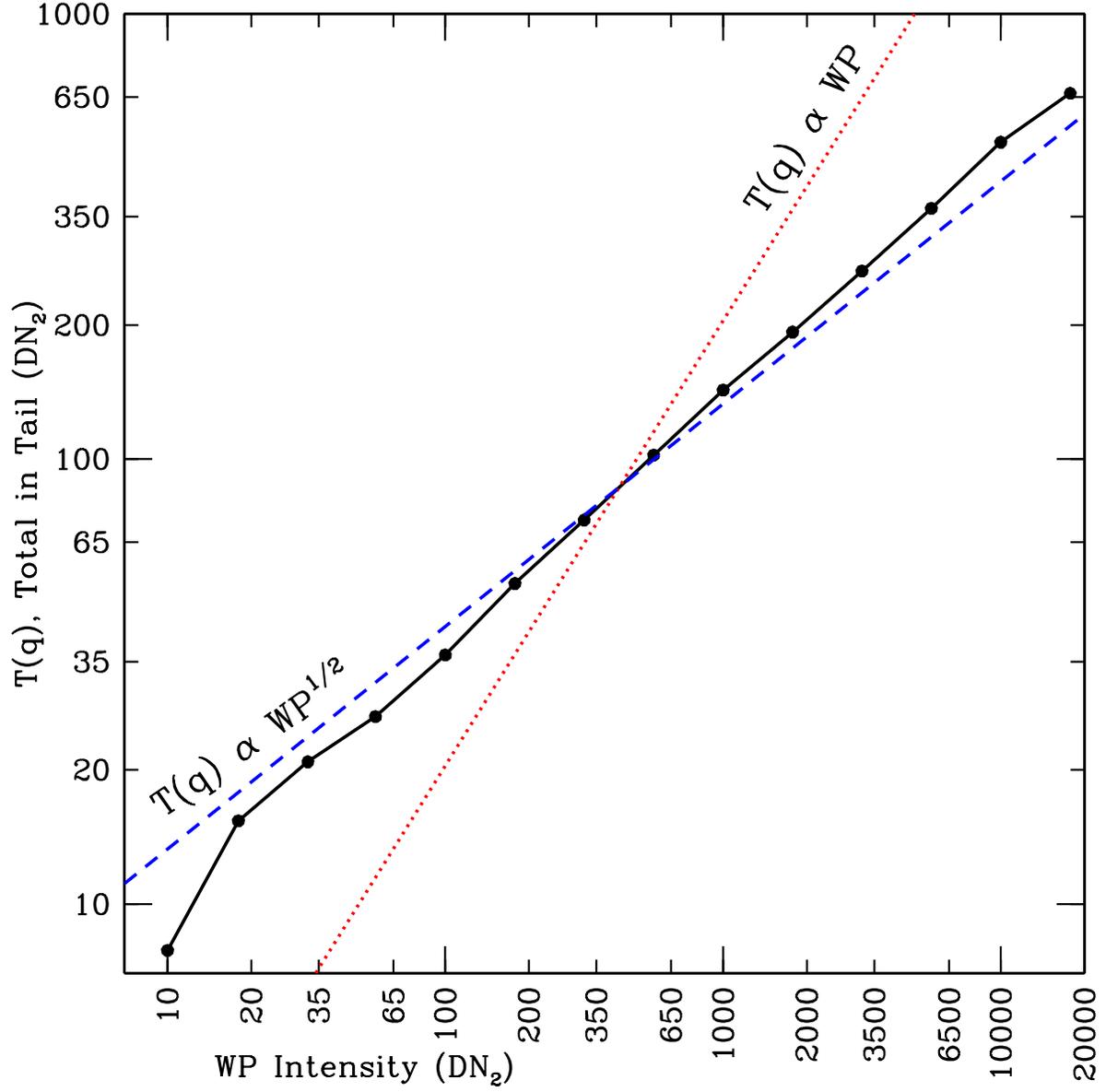}
\caption{We binned the profiles shown in Figure~\ref{fig05} more 
         finely and integrated up the flux in the profiles.  This 
         flux is plotted as a function of warm-pixel intensity, with 
         a log-log scale.  Lines with power-law slopes of 1.0 and 0.5 
         are drawn in, as labeled.
         \label{fig06}}
\end{figure}

In Figure~\ref{fig06} we integrate up the flux in the trails for 
profiles similar to those shown in Figure~\ref{fig05} (but with more
intermediate warm-pixel bins and plot the total amount of flux in 
the trail as a function of WP intensity.  The loss goes roughly 
as the square-root of the WP intensity, but not exactly.  We note
that the Philbrick (2001) model discussed in Bristow \& Alexov (2002) for 
STIS predicts that CTE losses should go as the square-root of the charge.
This rough power law is also seen in the experiments shown in 
Figure 10 of Janesick et al.\ (1991).  Hardy et al.\ (1998) 
simulate the distribution of charge set up within the pixel lattice
between the electrodes and show that packets with a more charge 
occupy a larger volume, and as such are subject to more traps.  
The relation they find in their simulations is very similar to this 
power-law scaling, though the details of any relation will no doubt 
depend somewhat on the particular pixel geometry and voltage settings.

\subsection{Instant or gradual shadowing?}
\label{ss.MODEL_SHADOW}

Before we can correct images for imperfect CTE, we must 
construct a model for what the readout process does to the 
original pixel distribution.  Modeling this process involves 
understanding in detail exactly when charge gets stuck in 
traps, and when it gets released.  Lab experiments and theoretical
models (see Cawley et al.\ 2001) predict that the probability 
of filling a trap is related to the local density of charge, 
with small packets taking longer to fill traps.  However,
given the slow parallel clocking speed, even small packets
are expected to fill all the open traps that they can access
almost instantly.  Since our aim is to model the readout process 
on a pixel-by-pixel basis, it will matter considerably to our 
model whether traps are filled instantly, or whether it may 
take several traversing packets to fill them completely.  In 
this section, we will use CRs in the individual dark images to 
evaluate the capture-time constant empirically.


If all the accessible traps do not get filled the instant a 
packet passes through, then we would expect that an object with 
two bright pixels would experience (and fill) more traps than 
one with only one bright pixel.  To test this, we compare the 
profiles upstream of WPs and CRs.
The warm pixels we studied above tend to have single high pixels, 
surrounded by a large number of low pixels.  By contrast, CRs tend to 
come in clumps of 8 or so impacted pixels.  We analyzed each of the 
168 images individually and identified bright CRs in the region at
least 1500 pixels from the serial register, where CTE losses would 
be largest.  We specifically selected CRs that had two bright pixels 
in the same column with nearly the same number of counts in each 
pixel to within 10\%, with the brighter pixel being farther from the 
register.  We further required that the CR be localized, such that 
the $\pm$ 5 pixels upstream of each CR had less than 5\% of 
the total of the flux in the two bright pixels.  This way, the CR 
should not interfere too much with the trail.

This experiment should allow us to assess whether the trail observed 
is more due to the sum of the pixels in an object 
(which would indicate a partial filling of the traps by the
first pixel), or whether all the accessible traps are filled 
completely the moment the first charge packet passes through 
(in which case the trail would reflect only the bright pixel 
that passed over).

\begin{figure}
\plotone{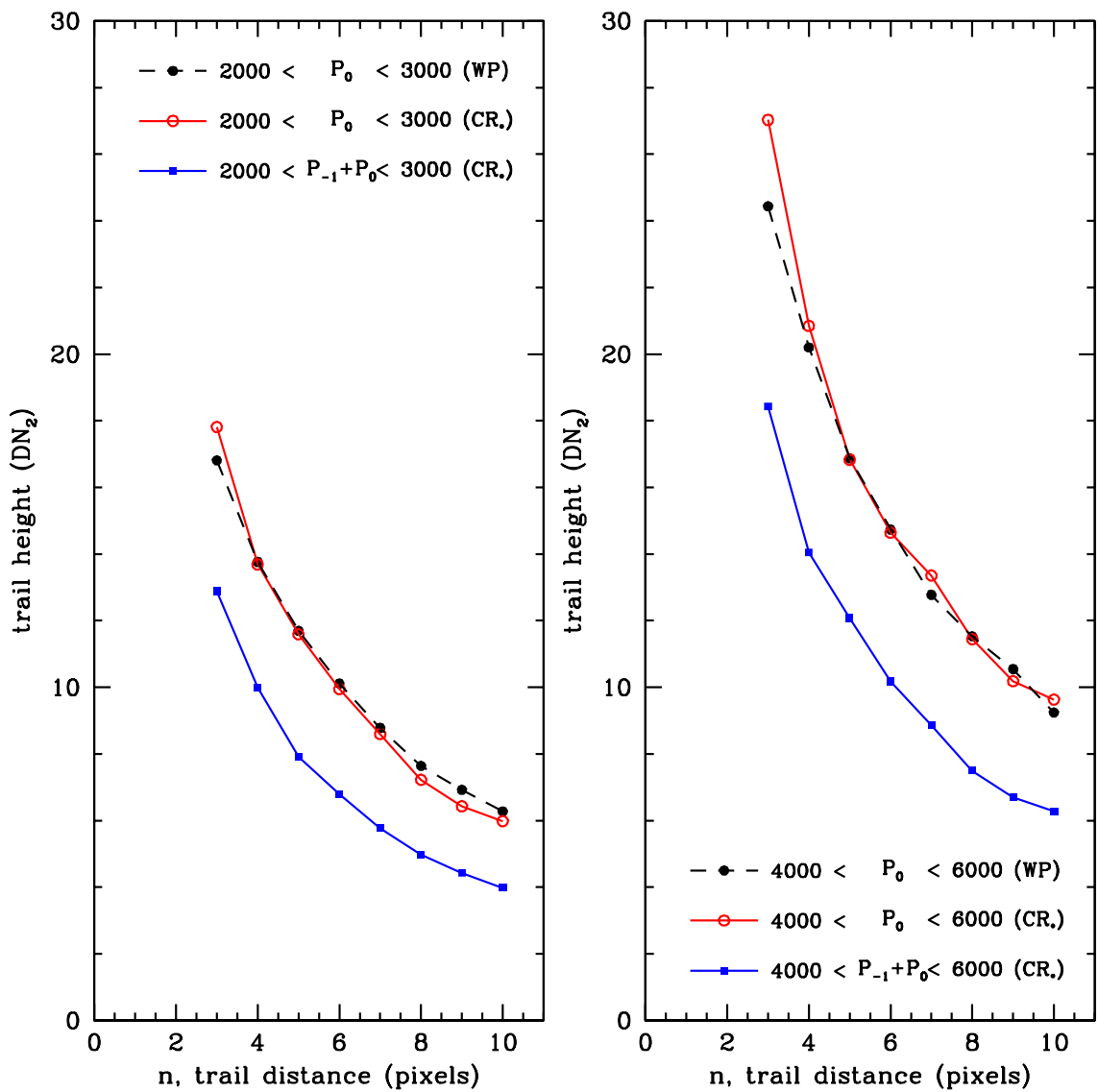}
\caption{We compare the CTE-related trails for CRs chosen to have 
         two nearly equal salient pixels against the trails for warm 
         pixels (which have one salient pixel).  CRs and WPs in the
         left panel have about 2500 DN$_2$, and those shown in the
         right panel have about 5000 DN$_2$.  The trail that is 
         observed clearly is regulated by the flux in the brightest 
         pixel, and not the total CR flux.
         \label{fig07}}
\end{figure}

Figure~\ref{fig07} compares the trails from WPs against those for 
the CRs, again for events at least 1500 pixels from the readout 
register.  The left panel shows in black the trails for WPs with 
intensities between 2000 and 3000 DN$_2$, and the right panel 
shows the same for WPs between 4000 and 6000 DN$_2$.  The red curve 
shows the trails for CRs where the brightest pixel falls within 
the same DN$_2$ range.  The blue curve shows the trails for CRs 
where the {\it total\,} of the two bright pixels falls within 
the same range as the WPs.  We show the trails for $n \ge 3$ 
only because even our selections could not guarantee that there 
would be no flux from the CR in neighboring pixels.
(The closest few pixels to the CR maximum do tend to have the most 
contamination, but the contamination does not always go immediately 
to zero.)

It is clear that the red profiles match the warm-pixel trail 
almost perfectly, while the blue profiles are markedly lower.  
{\bf This demonstrates that it is the flux in the brightest pixel, 
not the total flux, that regulates the profile of the trail.}  
We thus conclude that each pixel completely fills all the traps 
that are accessible to its electron packet, a conclusion that 
is consistent with the fast-fill assumption in Bristow (2002, p8).
This means that while the time to release an electron from a 
trap can be several parallel-shift times (see Figure~\ref{fig05}), 
the time it takes to fill a trap is negligibly small.  This 
characteristic should make the charge-transfer process much easier 
to model, as it means that we will not have to fold into our readout 
model an absorption probability, which would be extremely difficult 
to constrain.

\subsection{Statement of the model}
\label{ss.MODEL_STATEMENT}

Before we can remove the CTE effects from the observed WFC 
images, we must come up with a way to simulate what charge-transfer 
inefficiency does to observations.  The model we will construct 
below is similar in many respects to the one constructed in
Bristow et al.\ (2003a) and Massey et al.\ (2010).  Many of the 
limitations of these models stemmed from insufficient constraints 
from theory and laboratory experimentation.  In an effort to side-step 
these limitations, we adopt here a purely empirical formulation.  
Massey et al.\ developed their model using COSMOS images with a 
background of 51$\pm$9 $e^{-}$ ($\sim 25 {\rm DN}_2$).  As such, 
they were unable to probe CTE losses below about 100 $e^{-}$.  
Since it is worthwhile to probe the trap density all the way down 
to zero background, we will do our analysis on the dark-exposure 
images.  Only later will we make use of on-sky images to test the 
algorithm.  We provide a more detailed comparison between the assumptions 
of our model and the recent Massey model in \S\ref{ss.MODEL_MASSEY}.

Our model is quite simple.  When a packet of charge is transferred
through pixel B from pixel A on its way to the serial register, 
four things happen.  The first is that the untrapped flux in Pixel B 
is transferred out, either to Pixel C or to the serial-readout 
register.  The second is that the traps in pixel B that were filled 
previously release some of their counts.  Third, the packet of 
charge from pixel A is transferred into pixel B, and the recently 
released counts in the pixel are added to it.  The fourth thing 
that happens is that once this new packet is present in the pixel, 
it fills all the traps that it has access to.  At this point, the 
untrapped charge in the packet is ready to be shifted to 
the next pixel.  The number of counts that get transferred to the 
next pixel is number of counts in pixel A, {\it plus\,} the 
number of counts released by the traps in pixel B, {\it minus\,} the 
amount of charge that was newly retained by the traps in pixel B: 
$N_B = N_A + N_{\rm released} - N_{\rm trapped}$.  The $N_{\rm trapped}$ 
counts will remain in pixel B until released later into a different 
packet, associated with a different read-out pixel.

From our experiment in \S\ref{ss.MODEL_SHADOW}, we determined that 
every electron that can occupy a trap, will occupy the trap instantly.  
This means that when a packet is in a pixel, it fills all the traps it 
has access to.  The number of traps available to a packet is clearly
a function of how many electrons are in it.  Figure~\ref{fig06} 
showed that the number of traps accessible to a packet increases 
roughly as the square root of the number of electrons in the packet.  
Although this $T(q)$ curve goes nearly as the 0.5-power, it is not 
exact, so we will represent this relationship empirically by tabulating 
its value at a set of node points (10, 30, 100, 300, 1000, 3000, 10000, 
and 30000), using interpolation in between.  We note here that the charge 
$q$ here is in units of DN$_2$.

In practice, we will use the differential form of this curve, the number 
of traps per marginal electron, $\phi(q)$ = ${dT\over{dq}}(q)$.  The 
quantity $T(q)$ plotted in Figure~\ref{fig06} corresponds to the total
number of traps seen after about 1750 pixel shifts (we examined
pixels between $j$=1500 and $j$=2000). 

Calculating the number of counts released is more complicated, 
since charge is captured instantly, but it is released gradually.  It 
appears from Figure~\ref{fig05} that the traps that affect smaller 
charge packets have somewhat steeper release profiles than the 
traps that affect only larger charge packets.  This means that we may 
need to use a two-dimensional function to describe the release of 
charge.  The function $\psi(n,q)$ represents the probability that a trap
that affects the $q^{\rm th}$ charge will release its charge after 
the $n^{\rm th}$ pixel shift.  The distributions from Figure~\ref{fig05} 
should give us a rough idea of the shape of $\psi(n,q)$ for various 
values of $q$.

The one aspect remaining is to remember that electrons are conserved.
When an electron encounters a trap of a certain depth, there is a
probability, based on the distribution of electron packets that have
come before, that this trap will already be occupied.  If we later 
transfer charge into this pixel that would like to fill the
trap, we must remember that we can fill it only if it is empty.  We
can use the release-probability function $\psi(n,q)$ to tell us the
probability that the trap will be empty, and can then fill it with a
partial election, reflecting the expectation probability of an empty
trap.  Hardy et al.\ (1998) suggest a similar algorithm.

This begs an interesting question: will the trap then release
the electron according to when it was initially filled, or according
to when it was last filled (or ``topped off'')?  We will assume here
that when we fill a partially empty trap, we ``reset'' its
release-time constant, so the only thing that matters is when it was
last topped off.  This is the simplest possible assumption, and it
appears to be consistent with the data.
 
Even though all of the trap-filling and trap-releasing is surely a
quantum process involving one electron at a time, for simplicity 
here we treat it as a continuum process, dealing with fractional 
traps and fractional charge.  It is also true that each pixel has 
a different radiation-damage history and a different distribution of 
traps.  However, since there is currently no way for us to determine 
which pixels may hold traps of which depth (the density of traps is 
too high for pocket-pumping to yield a definitive distribution), 
it is reasonable to treat all pixels as having the same distribution 
of fractional traps.

\subsection{The Detailed Readout Algorithm}
\label{ss.MODEL_ALGO}

The readout simulator requires two functions that describe the nature 
of the traps:  $\phi_q$ and $\psi_{nq}$.  The $\phi_q$ array contains 
the number of traps from pixel {\tt j=2048} to {\tt j=1} that catch 
charge between $q - {1\over{2}}$ and $q + {1\over{2}}$.  For example, 
if a pixel at the top of the chip ($j=2048$) has 100 DN$_2$, then it 
will be subject to $\sum_{q=1}^{100} \phi_q $ traps.  Integrating up 
our final $\phi_q$ array gives a value of 36.3 DN$_2$ of traps (see 
$T(q)$ in Figure~\ref{fig06}).  This means that if the background is 
zero and a pixel near the top of the detector starts out with 100 DN$_2$, 
it will likely have 64 of them when it clocks out at the bottom.  
The trail will contain 36 DN$_2$.  This model only treats the charge 
that either stays with the pixel or ends up in the observable trail.  
We will see in Section~\ref{s.STARS} whether there are losses that do
not show up in the observed trails.

The second function is $\psi_{nq}$.  It gives the probability that 
the $q^{\rm th}$ charge in a packet will be released $n$ pixels 
downstream. The model we adopt is able to track charge trails out to 
100 pixels ($n=100$).  As we mentioned above, our model is not able 
to account for charge in the trails beyond this.  It is possible that 
there are even fainter trails that extend much further, but if they 
are present, we will have to use a different formalism to represent them.

For each column that is read out, there are 2048 $\times$ (2048/2)  
$\sim 2 \times 10^6$ total pixel shifts involved.  (The division by
two comes from the fact that a pixel in the top row gets shifted 
2048 times, and one in the bottom row only once; the typical row is 
therefore shifted 1024 times.) It would be computationally prohibitive 
to simulate each of these shifts individually for each of the 4096 
columns for each chip, so we recognized that many of the shifts are 
identical.  If we consider pixels A and B in rows $j=1001$ and $j=1000$, 
respectively, pretty much the same charge transfer takes place 1000 
times:  electron packet B is shifted out of a pixel, and packet A
is shifted in, and if CTE is close to unity, A and B will not change 
very much from the top to the bottom.  For simplicity, we model this 
as incurring 1000$\times$ the CTE loss as a single-pixel transfer.  
This simplification increases the algorithm's speed by a factor of 
$\sim$1000, and allows us to treat all pixels as having the same 
distribution of traps from $q$ = 0 to saturation.  Previous efforts 
(Massey et al.\ 2010 and Bristow et al.\ 2002) have assumed a specific 
realization of the distribution of traps throughout the chip in an
effort to make the calculation computationally tractable.

Our assumption above breaks down when CTE losses begin to have a 
significant impact on the charge distribution, such that we might 
expect different transfer effects at the top of the column, where the 
intrinsic structure is coherent, and at the bottom of the column, where 
the true structure may be broadened by imperfect CTE.  We recognize 
this potential limitation of our model, but such a situation should 
occur only in the more pathological situation where CTE cannot be 
considered a small perturbation.  For computational speed in 
developing and converging upon our model, we will treat all pixel 
shifts the same, but will be careful to identify situations where 
this simplified treatment may be inadequate (see 
Section~\ref{ss.SUMMARY_TODO}).

We simulate the column readout going from pixel {\tt j}$=1$ to pixel 
{\tt j}$=2048$ (or 2068 if we are including the virtual overscan).  
In order to monitor how long it has been since traps that affect 
different charge levels have been released, we maintain an internal 
array, $n_q$, which keeps track of how long it has been since the trap 
at each marginal charge level $q$ was filled.  At the beginning of the 
readout, the $n_q$ array is set to 100 for all $q$, from 1 to 50,000.  
This corresponds to all the traps starting out completely empty.  
As far as the model is concerned, it means they were last filled more 
than 100 transfers ago.

We now have all the elements necessary to illustrate the algorithm.  
We start by shifting the first pixel into the serial register.  At 
the beginning of the readout, this pixel has $P_1$ DN$_2$ in it.  
We use the following equation to determine how many electrons the 
traps in this pixel will hold back:
\begin{equation}
      {\tt N}_{\rm trapped} = \sum_{{\tt q=1}}^{P_1} \phi_q 
                                      \times
                              \biggl( 1 - \sum_{n=n_q}^{100} 
                                       \psi_{n_q q}
                              \biggr) \times 
                              \biggl( { j \over{2048}} \biggr).
\end{equation}
The first term within the main summation provides the number of 
traps present per 2048 pixels at each marginal charge level $q$.  
The second term (in parentheses) reports what fraction of the traps 
at this charge level are still unfilled.  The final term scales 
${\tt N}_{\rm trapped}$ for the number of pixel shifts to the register 
(just $j$).  The summation itself is over all the charge in the
pixel, $P_1$.

Finally, we shift ${P_1}^{\prime}=\bigl( P_1-N_{\rm trapped} \bigr)$ 
charges into the serial register, and this is what is read out by
the amplifier (modulo losses from CTE in the serial direction, which 
we will discuss briefly in \S\ref{ss.IMPR_XCTE}).  We then reset 
the $n_q$ array to 0 for all $q  \leq P_1$, since the traps were 
just filled all the traps up to this charge level.  

The next step is to determine how much charge is released by traps
in the pixel before the next charge packet arrives.  We do this by 
simply summing over the released charge, trap by trap:
\begin{equation}
     N_{\rm released} = \sum_{q=1}^{50,000} 
                            \phi_q
                            \times 
                            \psi_{n_q q}
                            \times 
                        \biggl( { j \over{2048}} \biggr).
\end{equation}
The summation is over all charges, from the ground to saturation.
The first term scales the release for the number of traps that affect
electrons at this charge level.  The second term corresponds to the
fraction of the charge in the trap that gets released in the 
${n_q}^{\rm th}$ transfer since the trap was filled.  The final 
scaling corresponds to the number of transfers that take place, with 
a pixel at the top of the column getting a factor of 1.00.  

Once the appropriate amount of charge has been released from each 
trap, we increment the counter for the trap at each charge level 
by setting $n_q$ to $n_q + 1$.  We cap $n_q$ at 100, since our model 
is unable to follow the trails beyond 100 pixels, and assumes that 
after 100 shifts, the traps are empty.  

Finally, we shift the packet of electrons from pixel $j=2$ into this first 
pixel and end up with $\bigl( P_2^{\prime} + N_{\rm released} \bigr) $ 
charge in the pixel.  At this point, the algorithm returns to the 
starting point above, and we determine how many electrons get captured 
in traps and transfer the remainder into the serial register.  This 
process continues for the pixels from $j=1$ to 2048 (or 2068 in the 
{\tt raw} frames).  

All of the dark exposures we have used here were taken at about the 
same time (within a two-month period in late 2009), so each exposure 
should suffer about the same CTE losses.  It is well established 
(see Riess \& Mack 2004 and Chiaberge et al.\ 2009) that CTE losses 
have been increasing linearly over time since ACS was installed in 
space during Servicing Mission 3B in 2002.  For times before or after
October 2009, we will need an additional multiplicative term to 
account for the overall scaling of trap density.  This term can go 
after the $\bigl( { j \over{2048}} \bigr)$ terms in the 
$N_{\rm trapped}$ and $N_{\rm released}$ equations above:
\begin{equation} 
  \biggl( { JD_{\rm OBS}                       - JD_{\rm SM-3B} \over{
            JD_{\rm 10\,\mbox{-}1\mbox{-}2009} - JD_{\rm SM-3B}}}
  \biggr).
\end{equation}
Since this model was developed for the October 2009 epoch, the
correction factor is naturally 1.00 for the present epoch.  

The adoption of the simple scaling law above implicitly assumes that 
nothing has changed in the detector's CTE properties from installation
to after SM4.  It remains to be seen whether anneals or changes
of CCD temperature have any affect on CTE trail profiles or scaling.
Laboratory experimentation and theory suggest that raising the
temperature should lower the release time and thus steepen the 
trails.  There is no reason to think that the number of traps will 
change within the $-77^{\circ}$ C to $-81^{\circ}$ C operating-temperature 
range.  All these issues should be thoroughly investigated before 
any general, long-term implementation of the algorithm.

\subsection{An Example}
\label{ss.MODEL_WALKTHRU}

\begin{figure}
\plotone{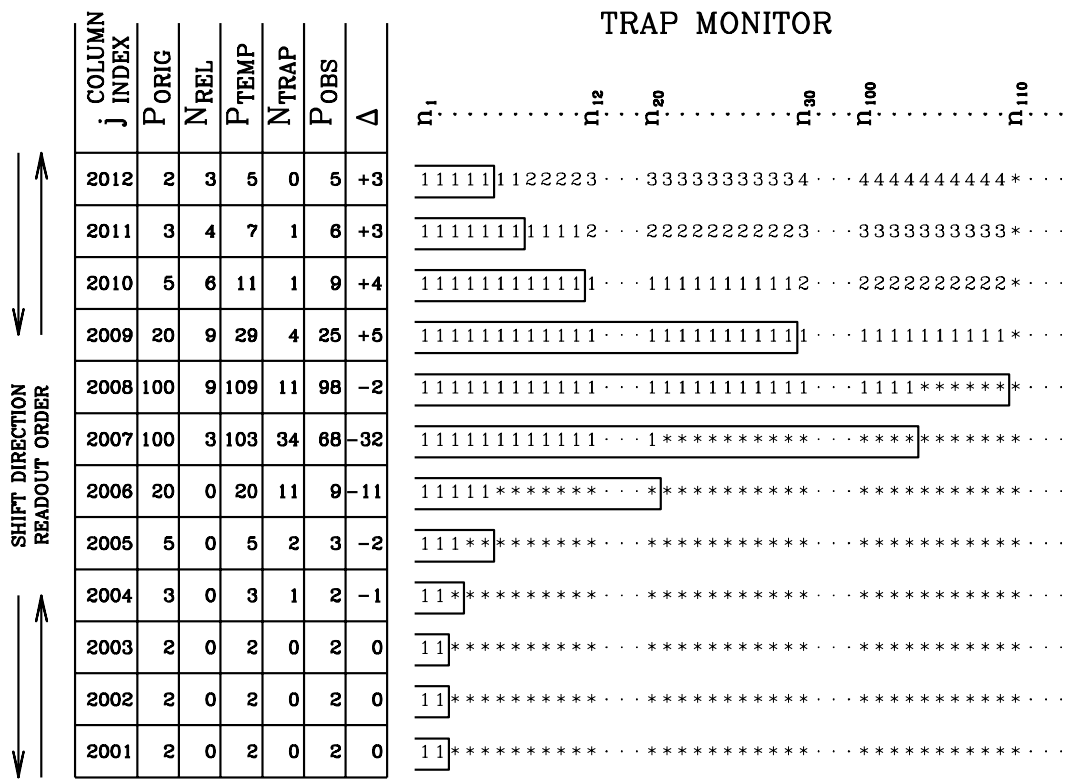}
\caption{This schematic shows the inner-workings of our 
         pixel-transfer
         algorithm.  See the text in 
         \S\ref{ss.MODEL_WALKTHRU} for details.  The asterisks
         in the trap monitor correspond to traps that are 
         completely empty ($n_q \ge 100$).  Numbers in the
         table are rounded to the nearest integer.
         \label{fig08}}
\end{figure}

Figure~\ref{fig08} provides an example of the readout-simulating
algorithm at work.  On the left, we show the original pixel value 
($P_{\rm ORIG}$) for pixels between column index {\tt j=2001} and 
{\tt j=2012}.  We simulate here a star that has its $y$-center at 
2007.5, with a background of 2 DN$_2$.
 
The next column in the table, $N_{\rm REL}$ shows the amount of 
charge released due to previous charge-filling from downstream pixels,
from Equation [2] above.  This charge is added to that in the previous 
column to get the third column, $P_{\rm TEMP}$.  This charge fills all
the holes available, which we compute from Equation [1] above.  The
trapped amount of charge is reported in the fourth column, $N_{\rm TRAP}$.
Finally, this is subtracted from the third column, to get $P_{\rm OBS}$, 
the charge that ultimately makes it to the serial register.  The sixth 
column $\Delta$ reports the difference between the input and output 
charge, $P_{\rm ORIG} - P_{\rm OBS}$.

The raster on the right reports the state of the charge traps.  The 
horizontal dimension corresponds to the marginal charge ($q$) that 
the trap impacts.  The number in this array ($n_q$) for each pixel 
at each charge level tells us how many transfers it has been 
since each trap was filled.  An asterisk here means that the trap 
was filled more than 100 pixels ago, and it is thus empty.  A {\tt 1} 
indicates it was filled by the previous pixel, a {\tt 2} by the pixel 
before that, etc.  The traps within the boxed region correspond to 
traps that will be filled by the current pixel (regulated by 
$P_{\rm TEMP}$), so that for the next upstream pixel, these traps 
will begin releasing with $n_q$=1.

It is clear in this simulation that the centroid of the star moves
considerably up the chip.  Rather than observing the true vertical 
profile of {\tt \{20,100,100,20\} }, we end up observing a profile 
of {\tt \{9,68,98,25\} }.  Note also that the first bright pixel 
ends up taking a large hit from imperfect CTE, losing 32 out of 
100 counts, but the second one loses only 2 out of 100 counts.  This 
is a result of the shadowing prescription implicit in our model (see 
\S\ref{ss.MODEL_SHADOW}).

\subsection{Iterating to Find the Original Pixel Distribution}
\label{ss.MODEL_ITER}

The above process tells us only what happens to an original distribution 
of charge in a column,  $P_j$, as it is transferred down the column 
and read out as ${P_j}^{\prime}$.  It does not tell us how to 
reconstruct the original ``source'' pixel array, $P_j$, from the
observed array.  To do this, we must somehow invert the process.  
We do this in a manner similar to that in Bristow \& Alexov (2002) and
Massey et al.\ (2010) (see their Section 3.2).

We first adopt the actual observed pixel array $P_{\rm OBS}$ as an 
estimate of the original array, $\hat{P}_{\rm ORIG}$.  We run this 
through the readout simulation and arrive at a simulated observed array,
$P_{\rm OUT}$.  This array will be less sharp than the actual observed 
array, since the readout process introduces CTE trails.  Our goal is 
to find the source array that, when run through the readout process, will 
yield the observed array $P_{\rm OBS}$.  So, we adjust the source array 
in such a way as to bring the output array to be closer to the observed 
array:  we add to the current source array the difference between 
the observed array and the output array.  The following pseudo-code 
shows how the convergence proceeds:
\newpage
\begin{tabbing}
\\
\,\,\,\,\,\,\,\,\,\,\, 
       \= $ {\bf \hat{P}_{\rm ORIG}^{[0]} }$ = $\bf P_{\rm OBS}$ \\
             \>   {\tt do } \= {\tt n = 1, 5 } \\
             \> \>   ${\bf P_{\rm OUT}^{[n]}}$ 
                     \= = {\tt COL\_READOUT[}
                     ${\bf \hat{P}_{\rm ORIG}^{[n-1]}}$ {\tt \,]} \\
             \> \>   ${\bf \hat{P}_{\rm ORIG}^{[n]}} $ 
                   \> $ = {\bf \hat{P}_{\rm ORIG}^{[n-1]}}   
                      + \biggl({\bf P_{\rm OBS}}
                             - {\bf {P}_{\rm OUT}^{[n]}}
                        \biggr)$  \\
             \> \>   {\tt enddo}. 
\end{tabbing}
The routine {\tt COL\_READOUT} takes in an array representing the 
original pixel values and simulates the charge-transfer process, returning 
the array that would be observed.  In the end, $\hat{P}_{\rm ORIG}^{[5]}$ 
is our 5$^{\rm th}$-iteration estimate of the original pixel 
distribution.  Most charge, even when the background is low, comes 
through the read-out process without being shifted, so imperfect CTE 
can be thought of as a slight perturbation.  As such, this process 
converges quite quickly.  We generally do 5 iterations.

One slight concern about this iterative scheme is that the flux recorded
in each image pixel ${\bf P_{\rm OBS}}$ does not correspond exactly 
to the amount of charge in DN$_2$ that arrived at the register.  The 
observed pixel value also contains a contribution from readnoise 
($\sim$4.5 $e^{-}$ for ACS's WFC).  If we operate the algorithm on 
${\bf P}_{\rm OBS}$, then the process will end up amplifying the readnoise, 
which may not be desired.  In \S\ref{ss.IMPR_RNMIT} we explore one way 
to minimize the influence of readnoise-level fluctuations on our algorithm.
However, it should be safe to defer the readnoise-related issues until 
later, as our dark exposures have been combined into a stack that 
should have the noise reduced by $\sqrt{168}$.

\subsection{Finding the optimal model parameters}
\label{ss.MODEL_PARAM}

To keep our model as simple as possible, we wanted to use the minimum
number of parameters.  The full model as stated would have about 
5 million degrees of freedom if we consider all the possible values 
of $\phi_q$ and $\psi_{n,q}$ for all $q$ and $n$.  Clearly these functions 
will vary smoothly with their parameters and we can represent them 
interpolating an adequately spaced table.  To this end, we tabulate the 
value for $\phi_q$ at an array of logarithmically spaced points in $q$: 
{\tt 1,3,10,30,100,300, 1000,3000,10000,30000,} and {\tt 100000} DN$_2$, 
and use log-linear interpolation to compute $\phi_q$ for any integer
$q$.  Similarly, we tabulate the two dimensional function $\psi_{nq}$ 
at values for $n$ shifts of {\tt 1, 2, 3, 5, 8, 12, 16, 20, 25, 30, 
40, 50, 60, 70, 80, 90,} and {\tt 100} pixels, and values of $q$ of 
{\tt 10, 100, 1000,} and {\tt 10000} DN$_2$.  To get $\psi_{nq}$ 
for any value of $n$ and $q$, we interpolate linearly in the $n$ dimension 
and logarithmically in the $q$ dimension.  This interpolation is done 
once-for-all, so that the program ends up working with arrays 
{\tt phi[1:100000]} and {\tt psi[1:100,1:100000]}.  The condensed set 
of parameters can be conceptualized as having 79 nodes:  $\Phi$[1:11]
and $\Psi$[1:17,1:4] (the capital letters denote the node parametrization).

Unfortunately, we cannot simply take the extracted values for these
functions from Figures~\ref{fig05} and \ref{fig06} to arrive at the 
final model.  The trail for a bright pixel generates many lower-intensity 
pixels in its lee.  The charge released in the trail pixels is subject 
to the same imperfect transfer as the original charge, so there is 
some interplay between the bright and faint trails.  

We resolved this complication by iteration.  We started with the 
arrays as extracted directly from Figures~\ref{fig05} and \ref{fig06}.  
We then ran the reconstruction routine on the dark stack to see how well 
the trails were removed.  We again constructed a plot similar to
Figure~\ref{fig05}, only this time with the reconstructed image as 
the source image.  The remaining trends showed us how to adjust 
$\Phi_Q$ and $\Psi_{NQ}$ to in such a way as to improve the fit.

It took several hands-on iterations for the model parameters to 
converge.  We began the convergence with traps at all $q$ levels 
having the same trail profile (i.e., $\psi_{nq}$ being a function 
of $n$ only).  This worked well for all the traps above $q \sim 100$ 
DN$_2$, and it implies that the slightly different profiles seen in 
Figure~\ref{fig05} may be a result of a some self-shielding within 
the trails.  It also means that most traps, irrespective their location 
within the pixels, appear to have largely the same release-time 
profiles.  

We were unable to fit the very lowest bin ($q \sim$ 10 DN$_2$) 
with the same profile, so we gave this lowest bin a steeper profile.
Even with a steeper profile, we had to use an extremely high trap
density (0.8) to account for the amount of charge in these trails.
This means that 80\% of the electrons at this charge level are likely 
to encounter a trap when transported from the top of the chip to the 
bottom.  This implies a significant level of CTE losses, and a more 
sophisticated, non-perturbative approach may be necessary to deal 
properly with CTE issues at this low flux level (see the discussions in 
\S\ref{ss.MODEL_ALGO} and \S\ref{ss.SUMMARY_TODO}).

Once CTE losses reach a truly pathological level, it will no longer be
possible to make any inference about the original charge distribution.
This will happen when there are more traps than charge, such that
most charge will be removed from its original pixel packet and can no
longer be associated with its original pixel location.  The deep 
dark images currently available may not be the best way to explore 
these issues, since in the 1000 s exposures, there are so many 
$\sim$10 DN$_2$ WPs, that it is hard to study them in isolation.  
Taking shorter darks should remedy this, but for the moment, we 
will simply adjust our model to deal with the low-$q$ WPs as well as 
possible, and will see how well the algorithm can work.  Improvements
can be made later (see \S\ref{s.SUMMARY}).

\begin{figure}
\plotone{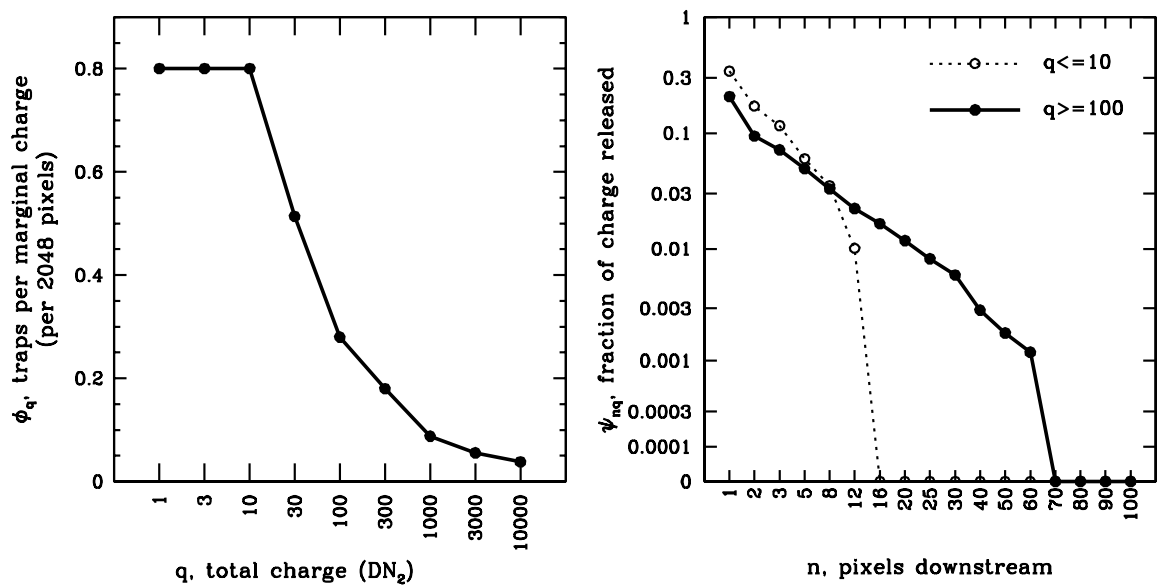}
\caption{This shows the run of parameters used in the converged-upon
         model.  On the left, we show $\phi_q$, the number of traps in
         a column per marginal charge at $q$.  On the right, we show
         the release profile for $q \leq 10$ (dotted curve) and for 
         $q \geq 100$ (solid curve).  The symbols indicate the location
         of the nodes used to specify the functions.
         \label{fig09}}
\end{figure}

The converged-upon model parameters are shown in Figure~\ref{fig09}.  
The trap density $\phi$-vs-$q$ is shown on the left.  The trend largely 
follows $\phi_q \sim q^{-0.5}$, consistent $\phi_q$ being the derivative 
of $T_q$ (see Figure~\ref{fig06}).  The release profiles are shown on 
the right, normalized to have a total probability of 1.00.  The steeper 
profile applies for $q \sim 10$, the other for $q \geq 100$.

\begin{figure}
\plotone{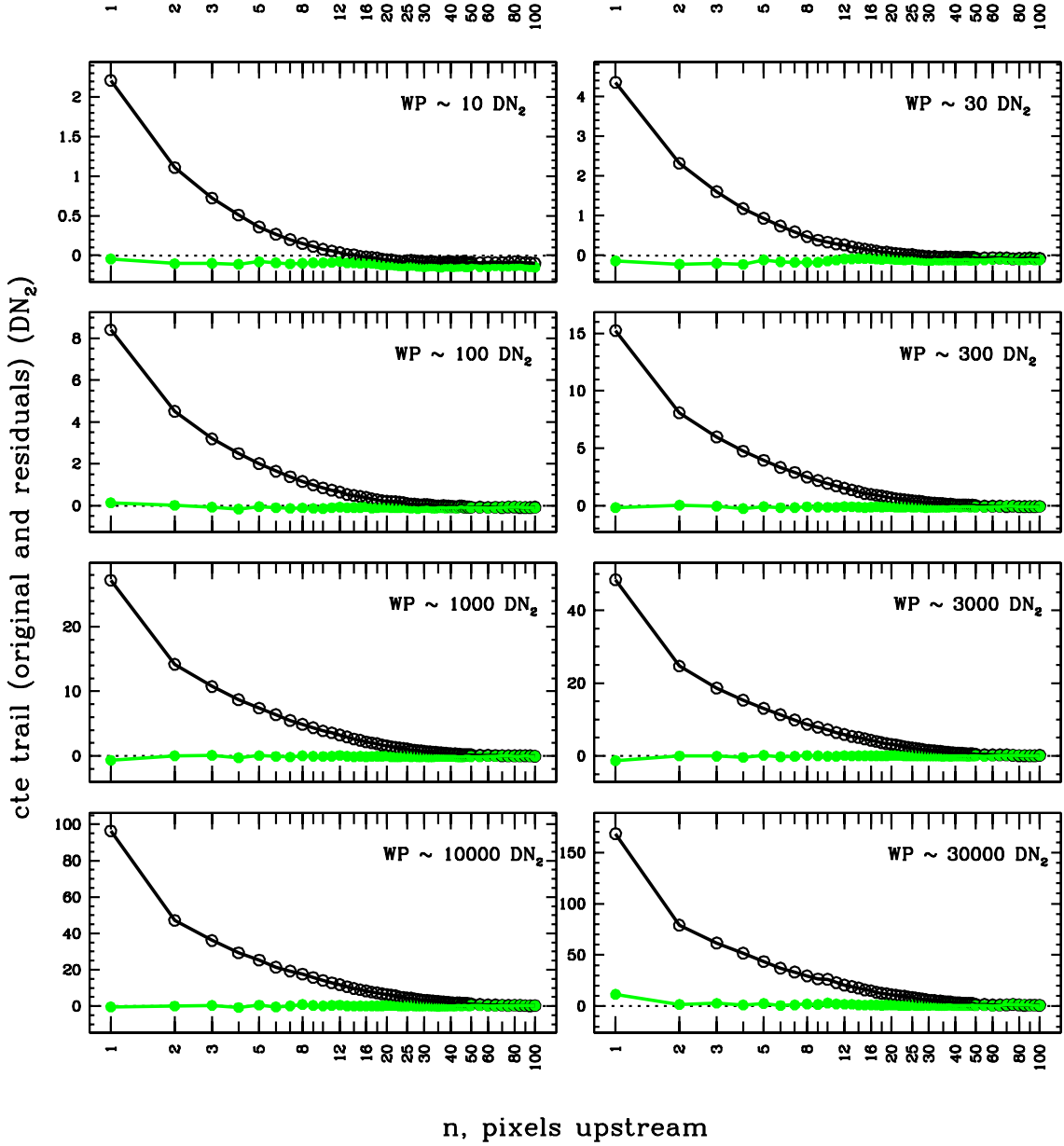}
\caption{Similar to Figure~\ref{fig05}.  The original trails (black) 
         and the residuals (green) after model-fitting for the
         warm-pixel stack, binned for 8 different warm-pixel intensity
         levels.
         \label{fig10}}
\end{figure}

The first real test, of course, is how well the trails are modeled.  
Figure~\ref{fig10} shows the residuals of the trails, in the same 
format as Figure~\ref{fig05}, though with twice as many warm-pixel 
bins.  It is clear that the trails for all levels of charge are 
removed cleanly.  The trails appear to be removed well even in the 
faintest WP bin, where our algorithm has to treat larger CTE losses 
than it is designed for.

\begin{figure}
\plotone{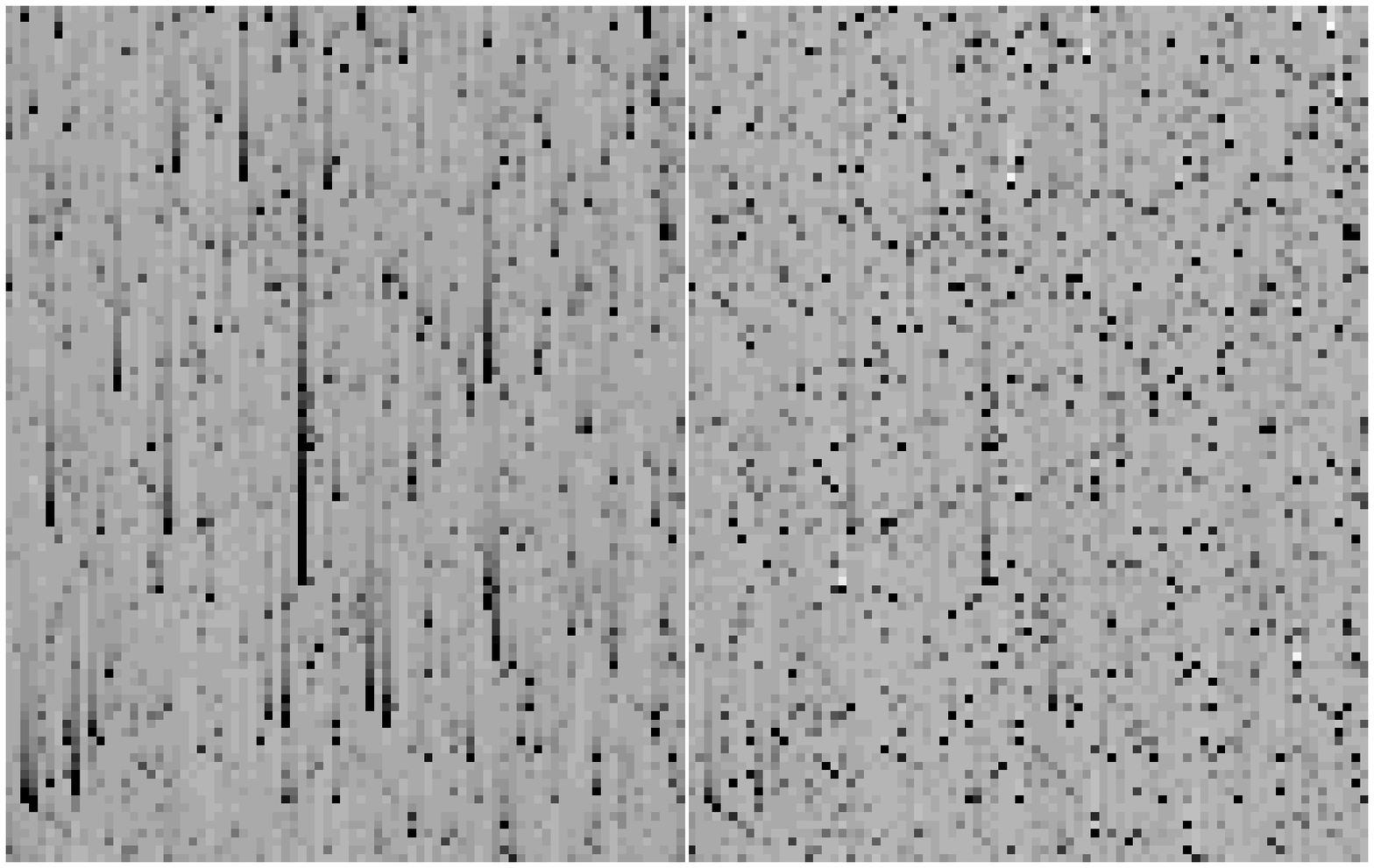}
\caption{(Left) The stack of the dark exposures, after the bias has 
         been removed.  (Right) The same stack, after the pixel-based 
         Y-CTE correction. 
         \label{fig11}}
\end{figure}

Figure~\ref{fig11} shows the before and after images centered on the
same region as in Fig.~\ref{fig03}.  It is clear that we are removing 
a large fraction of the trails and restoring the flux to its rightful
pixel --- usually a single pixel in these dark, warm-pixel pocked 
images.  Note that there is nothing in the reconstruction algorithm 
that would force flux back into one pixel, as opposed to two or three.
The restored image is simply the one that is best able to predict the 
observed image.  The fact in most cases that the flux was restored 
to a single pixel is an encouraging sign.  

There are clearly some trails in the image that are over-subtracted, 
and others that are under-subtracted.  It is interesting to note that 
the over-subtracted trails often lie in the same rows, as do the 
under-subtracted trails.  A significant WP of 1000 DN$_2$ loses about 
150 DN$_2$ to the trail (see Figure~\ref{fig06}).  If each trap impacts 
a single electron, this implies 300 traps (1 DN$_2$ = 2 e$^{-1}$).
If these traps are distributed randomly, we should expect a 
column-by-column variation in CTE losses of $\sim$5\%.  This 
variation would be even larger if the traps are generated in 
clumps, many-at-a-time.  Waczynski et al.\ (2001) note that 
energetic photons and electrons tend to generate single traps, 
but neutrons and ions often generate multiple traps in the same 
pixel, such that the particular radiation environment could make 
CTE more variable from column to column than single-trap Poisson 
statistics would suggest.  We can better study the statistics of 
trail subtraction in the individual exposures, using CRs or stars, 
as these (unlike the WPs) impact different places on the detector 
in different exposures

\subsection{Limitations of the super-dark analysis}
\label{ss.MODEL_LIM}

Since the dark frames have very low background and therefore suffer 
the largest CTE losses, it is extremely encouraging that our algorithm
does so well here.  Nevertheless, it will still be important to show that 
the parameters can work in ``easier'' (and more scientifically relevant) 
environments, as well.  Thus, we will have to test the algorithm on 
sky-exposed images that have a variety of background intensities.  

As far as what to expect with the higher background situation, we 
note that the model as formulated does naturally predict lower CTE losses 
when the background is higher, since empirically we see more traps that 
affect small charge packets than those that affect only large packets.
When the background is high, these small-packet traps are all filled 
and it is only the marginal larger-packet traps that come into play. 
 
One additional inadequacy of developing the correction on the stacked
dark frame is that there are not many high-signal warm pixels.  This
makes it difficult to calibrate the upper end of the $\phi_q$ curve.  
In the next section, we will examine the bright end more carefully 
using stars in short exposures.

\subsection{Final Model}
\label{ss.MODEL_FINAL}

The parameters of the converged-upon final model shown in 
Figure~\ref{fig09} are listed in Tables~\ref{tab02} and \ref{tab03}.  
These parameters, fed into the algorithm discussed in \S\ref{ss.MODEL_ALGO} 
result in a near-perfect removal of the CTE-related trails from 
the dark exposures.  It remains to be seen whether this restorative 
fix is sufficient to account for {\it all\,} CTE losses.  It is 
possible that the CTE trails could have a longer and fainter component 
than we can observe empirically.  The final test will therefore be to 
run this pixel-based correction on images for which we know what the true 
scene should look like.  This will be the focus of the next sections.


\begin{table}
\begin{center}
\caption{The number of traps per marginal DN$_2$,
         $\phi_q$, from the optimized model. 
         Recall that 1 DN$_2$ = 2 electrons.  \medskip }
\label{tab02}
\begin{tabular}{|r|r|r|}
\hline 
\multicolumn{1}{|c }{ node      } &
\multicolumn{1}{|c }{ marginal  } &
\multicolumn{1}{|c|}{ traps per } \\
\multicolumn{1}{|c }{ number    } &
\multicolumn{1}{|c }{ DN$_2$    } &
\multicolumn{1}{|c|}{ DN$_2$   } \\
\hline 
\multicolumn{1}{|c }{ $Q$  } &
\multicolumn{1}{|c }{ $q$  } &
\multicolumn{1}{|c|}{ $\phi_q$  } \\
\hline 
1  &       1 & 0.8000 \\
2  &       3 & 0.8000 \\
3  &      10 & 0.8000 \\
4  &      30 & 0.5000 \\
5  &     100 & 0.2800 \\
6  &     300 & 0.1750 \\
7  &   1,000 & 0.0880 \\
8  &   3,000 & 0.0540 \\
9  &  10,000 & 0.0380 \\
10 &  30,000 & 0.0115 \\
11 & 100,000 & 0.0060 \\
\hline 
\end{tabular}
\end{center}
\end{table}


\begin{table}
\begin{center}
\caption{The measured release profile for traps at different 
         $q$ levels ($q$ is in DN$_2$).  Note the model allows
         release times up to 100, but we found 
         no perceptible charge in the trails beyond $n$ = 70 shifts.
          \medskip}
\label{tab03}
\begin{tabular}{|r|r|rr|}
\hline 
\multicolumn{1}{|c }{node} &
\multicolumn{1}{|c }{shifts} &
\multicolumn{2}{|c|}{fractional release} \\
\multicolumn{1}{|c }{$N$} &
\multicolumn{1}{|c }{$n$} &
\multicolumn{1}{|c }{$\psi_{n,q \leq 10}$} &
\multicolumn{1}{ c|}{$\psi_{n,q \geq 100}$} \\
\hline 
1 &   1 & 0.3417 & 0.2061 \\
2 &   2 & 0.1709 & 0.0942 \\
3 &   3 & 0.1156 & 0.0718 \\
4 &   5 & 0.0603 & 0.0495 \\
5 &   8 & 0.0352 & 0.0330 \\
6 &  12 & 0.0101 & 0.0224 \\
7 &  16 & 0.0000 & 0.0165 \\
8 &  20 & 0.0000 & 0.0118 \\
9 &  25 & 0.0000 & 0.0082 \\
10&  30 & 0.0000 & 0.0059 \\
11&  40 & 0.0000 & 0.0029 \\
12&  50 & 0.0000 & 0.0018 \\
13&  60 & 0.0000 & 0.0012 \\
14&  70 & 0.0000 & 0.0000 \\
15&  80 & 0.0000 & 0.0000 \\
16&  90 & 0.0000 & 0.0000 \\
17& 100 & 0.0000 & 0.0000 \\
\hline 
\end{tabular}
\end{center}
\end{table}


\subsection{Comparison with the Massey Model}
\label{ss.MODEL_MASSEY}

Since there is already a pixel-based CTE correction for ACS's WFC that
has been presented in the literature and made available for users, it 
is worthwhile to note here the ways in which our approach is similar 
to and different from that of Massey et al.\ (2010).

Our approaches are similar in several respects.  First, both approaches
model the pixel-transfer process, keeping track of the number of 
electrons in each pixel's packet, and the number of electrons that get 
trapped and released in each transfer.  We both assume that a trap will 
grab an electron the instant the trap is open and has an accessible 
electron to trap.  We also both model the trap density as a function 
of packet intensity, finding that the cumulative number of relevant traps 
goes roughly as the square root of the amount of charge in the packet.

The first major difference between our models is in terms of the profile 
of the trails.  Rather than fit the trails with hard-coded dual exponentials, 
our model uses a purely empirical drop-off, parametrizing the 100-pixel 
trail with 17 nodes, spaced every pixel at smaller offsets and every 
ten pixels at larger offsets.  This flexibility allows us to more 
accurately tailor the drop-off to the observations.  Figure~\ref{fig05} 
shows that a dual exponential does not adequately describe the entire 
visible trail. 
 
Another difference is that our treatment deals naturally with 
variations in the background.  One of the variables they invoke in 
their \S\,2.5, $n_q(y,n_e,b)$, explicitly acknowledges that the background 
will impact the CTE losses.  However, since all of their development 
data were taken with backgrounds of $51 \pm 9$ electrons, they were 
unable to explore this component of their model.  They implemented
the idea of a ``notch'' at the bottom of the pixel well that lets 
through without loss any charge packet with $n_e < 100$ electrons.
Since our model was constrained using data with the lowest possible 
background, we were able to probe the trap density all the way 
down to the traps that affect even small electron packets.  We find 
that, rather than dropping to zero at $n_e < 100$ as their model 
stipulates, the trap density actually continues to increase below 
this fill-level\footnote{
     The WFC detectors were manufactured to have a ``notch'' or 
     ``mini-channel'', so that the smallest charge packets could
     be transferred with significantly less CTE loss.  Unfortunately,
     the doping did not stay in the small channel but dispersed 
     into the rest of the silicon lattice, thus resulting in no
     trap-free channel. Manufacturing procedures have improved considerably
     since the ACS chips were made.  (ACS Technical lead, D. Golimowski, 
     personal communication).
     }.  
The main impact of increasing the background is to 
pre-fill and take out-of-play the lowest charge traps.  As such, 
we anticipate that our algorithm should work with all kinds of 
backgrounds, not only those brighter than 51 electrons.  This 
claim will be explicitly evaluated in \S\ref{ss.MODEL_BKGD}.

The third major difference between our models concerns how traps are 
simulated.  Massey et al.\ (2010) simulated placing explicit traps 
that affect single electrons at random locations within the pixel 
grid.  While they experimented with making more traps that could 
affect partial electrons, or fewer traps that would each affect 
multiple electrons, at heart their treatment was a quantized 
treatment of the traps.  Each of their traps was also thresholded 
to affect electron packets at specific charge level.  Neither the 
placement of the traps nor their thresholds were empirically 
determined to match the detectors, but were simply one possible 
realization of the detailed detector based on their model.  Our 
treatment, on the other hand, deals with traps in a continuum 
sense.  Whereas they modeled about half the pixels as having a 
trap of one variety or another, we modeled {\it all\,} pixels as 
having an array of fractional traps that affect packets from 
$q$ = 1 to 40,000 DN$_2$.  Consistent with our continuous approach, 
we find for an observed 2-byte-integer pixel array with a quantized
number of electrons a 4-byte-real source array, which can have 
fractional electrons, and which essentially represents the 
expectation value for how many electrons were most likely present
before the charge-smearing readout process. 
 
A final difference is that we found the need to iterate more than once,
whereas they determined a single iteration to be sufficient.  We found 
that only after 3 or so iterations did the modeled and observed pixel 
arrays converge.  We generally iterate 5 times.  Perhaps our need for 
more iteration is related to the more continuous way we modeled the 
traps.  It could also be related to the fact that we are working with 
a stack of 168 images, and the noise threshold is much lower.

\subsection{Examining WPs on different backgrounds}
\label{ss.MODEL_BKGD}

The low background in the dark exposures allowed us to detect the 
very faint trails behind very low warm pixels.  We cannot, however, 
use these dark images to test the algorithm on more typical science 
images, which tend to have backgrounds between 20 and 150 electrons 
(10 and 75 DN$_2$).  The tests we ran with CRs (see Figure~\ref{fig07}) 
seem to validate the assumptions that our model makes in regards 
to shadowing, namely, that a trap will absorb an electron the instant 
it can.  Electrons in the sky background should work in a way similar 
to the leading pixels in the cosmic rays we studied:  they should 
pre-fill some traps, so that the upstream charge is subject only 
to the traps that are above the background level.  However, this 
needs to be empirically verified.

We examined three different F606W images taken of the 47 Tuc calibration
field with exposure times of 30 s, 150 s, and 350 s, and with backgrounds 
of 1.5, 8, and 16 DN$_2$.  We applied our CTE-restoration scheme to 
each of the {\tt \_raw} exposures to produce a corrected raw image and
used the warm-pixel locations identified from the peak-map of our 
previous study.  We then examined the WP profiles in the original and 
corrected images.  Since these results come from individual exposures, 
rather than the dark stack of 168 exposures, we would naturally expect 
things to be much noisier, even if the general trends are corrected.

\begin{figure}
\plotone{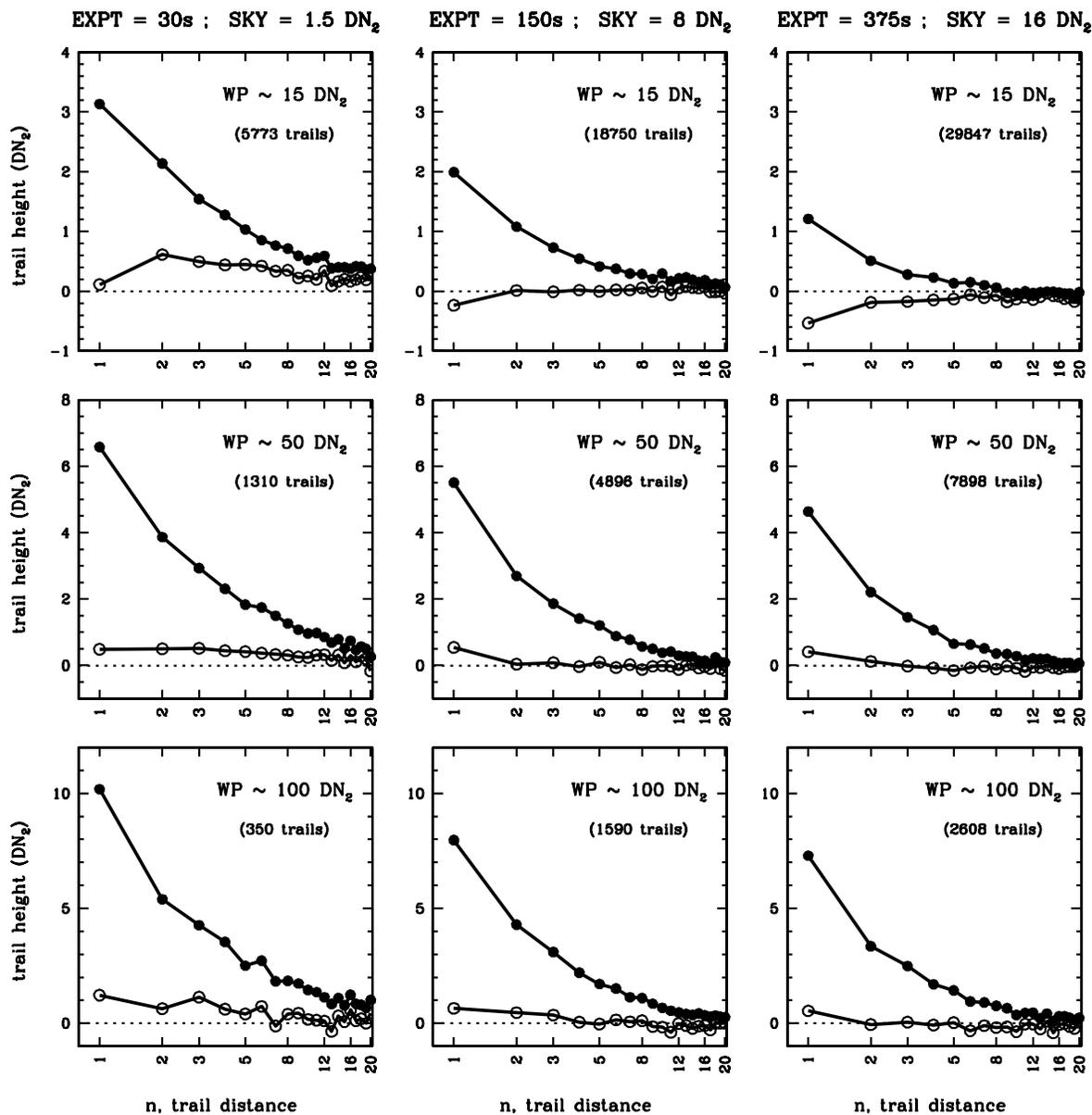}
\caption{The CTE trail profiles for warm pixels in the original 
         and corrected {\tt \_raw} images images of the 47 Tuc 
         calibration field.  The curve with the filled symbols 
         represents the raw, observed trails.  The open-symbol 
         curve represents the trails in the corrected exposures.
         The panels on the left show 30 s exposures with a background 
         of only 1.5 DN$_2$, the middle panels show the results 
         for 150 s exposures, with a background of 8 DN$_2$, and 
         the right panels show 375 s exposures with a background 
         of 16 DN$_2$.  In each column, the panels from top to 
         bottom show the results for faint to moderate warm pixels.
         \label{fig12}}
\end{figure}

Figure~\ref{fig12} shows the results in an array of panels, for 
warm pixels with intensities of 15, 50, and 100 DN.  We could not 
cross-compare brighter WPs, since the short 15 s exposures allowed 
few warm pixels to get very bright.  It is illustrative to compare the 
trends across rows.  On the left, where the background is lower, the 
observed trails are clearly larger than on the right, where the 
background is higher.  This effect is almost a factor of 3 for the 
faintest WPs, and a factor of 1.5 for WPs with $\sim$100 DN$_2$.

The corrections do a good job for all cases, naturally tracking the 
disparate CTE impact with background.  In most cases, the flux in 
the trail is reduced by well over 90\%.

This first set of tests on the background is encouraging.  It tells
us that our algorithm properly accounts for the shadowing that results
as the background pre-fills traps.  This, in turn, results in less 
CTE loss for faint objects.  The next tests will involve studying 
how the algorithm works for stars, which are more complicated than 
the delta-function-like WPs.  Also, since we know the true fluxes, 
positions, and shapes for stars, we can assess our correction in an 
absolute sense.

\clearpage


\section{TESTING WITH STARS}
\label{s.STARS}

The restoration of charge into the delta-function-like warm pixels 
is a basic test that any pixel-based algorithm should pass.  This 
test is a necessary---but not a sufficient---demonstration that the 
algorithm is robust for science images.  It is possible that there 
could be some species of traps that release the charge over much 
longer time-scales (and hence, longer trail distances) than can be 
measured with even bright sources.  Such traps would leave no 
perceptible trail, yet they could still remove significant charge 
from objects. 
 
The only way to test whether our algorithm actually does restore 
all the lost flux is by comparing various restored images against 
the truth.  This can be done in two ways.  In the first test, a 
shallow-exposure image can be corrected and compared against a 
corrected deep-exposure image.  Since the CTE losses for the the 
first exposure would be much larger than for the second (thanks 
to the higher background in the second), the comparison provides
a direct empirical test.  This comparison could involve extracted 
parameters such as fluxes or positions, or it could be made on a 
pixel-by-pixel basis.  The second test will involve comparing two 
images that are dithered vertically by a whole chip, which maximally 
enhances the CTE contrast.

Before embarking on these tests, however, we should perform 
the \ae sthetic test.  Does it {\it look\,} like the trails have 
been removed in science images?

\subsection{Removal of Trails from stars}
\label{ss.STARS_BYEYE}

\begin{figure}
\plotone{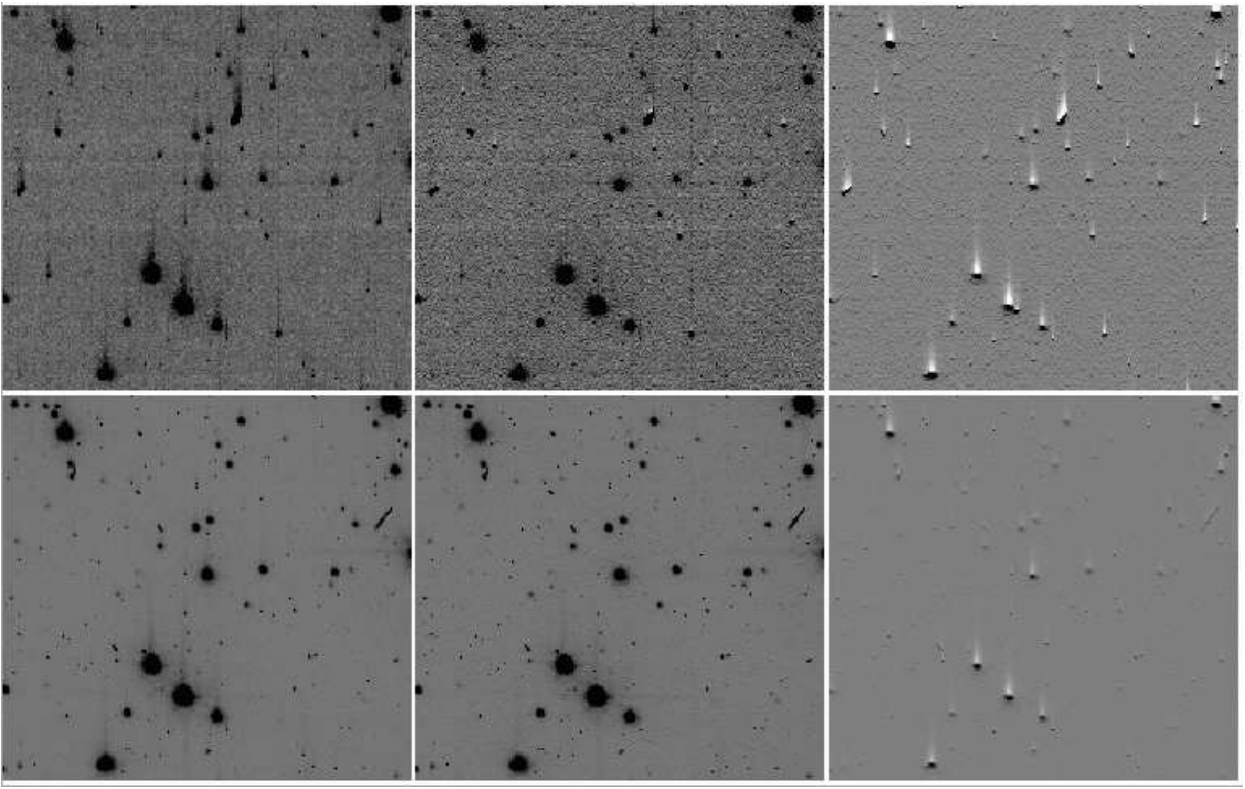}
\caption{Panels from left to right show, respectively, the original 
         {\tt raw} image, the corrected {\tt raw} image, and the 
         difference between the two.  The top panels show the results 
         for {\tt ja9bw2yq} a 30 s exposure, and the bottom panels 
         show them for {\tt ja9bw2lq}, a 349 s exposure.
         \label{fig13}}
\end{figure}

In the panels of Figure~\ref{fig13}, we show the uncorrected, corrected, 
and difference images for two exposures of the 47 Tuc calibration 
field:  {\tt ja9bw2ykq\_raw}, which had an exposure time of 30 s and 
a low background (1.5 DN$_2$, top) and {\tt ja9bw2ylq\_raw}, which 
had an exposure time of 349 s and a higher background (16 DN$_2$).  
In both cases, the CTE trails from stars are obvious in the original
images on the left, and are indiscernible in the corrected images
in the central panels.  The difference images show how the 
correction algorithm has redistributed the flux.

The scaling for the image panels has been matched to the exposure 
time, so that the more significant corrections in the upper-right 
panel, as compared against the lower-right panel, indicate a much 
larger fractional adjustment for the lower-background images.

It is worth noting that many of the CRs in the short exposure in the
top panels have their trails over-subtracted, while this does not appear
to be the case for the 349 s exposure.  The reason for this is that the 
detector is not read out instantaneously.  The read-out takes about 
100 s seconds (see \S\ref{ss.DARK_DATA}) and the pixels in these 
short exposures thus spend more time getting read out than they do 
exposing on-sky.  Consequently, it is more likely than not that an 
observed CR will have struck the detector during readout, rather than 
during the formal exposure time.  As such, this late-hitting CR will 
not have been transferred across the entire array, but only part of 
it, and the correction, which assumes $j$ pixel-shifts for an object 
appearing in row $j$, will naturally over-correct.

The fact that the corrections developed to remove the CTE trails from
warm pixels also appear to do a good job restoring flux to stars is 
a good sign that the algorithm has universal application.  The real 
test, however, will involve a more quantitative analysis.

\subsection{Testing the algorithm on {\tt \_flt} images}
\label{ss.STARS_FLT}

The raw corrections in the preceding sections were performed 
on the {\tt raw} images, which have not gone through the standard 
pipeline process of bias-subtraction and flat-fielding.  It is 
worth asking whether it would be safe to work with the {\tt flt} 
exposures, instead.  It would certainly be easier if the algorithm 
could operate on pipeline products, so that it would not be necessary 
to re-run the pipeline calibrations by hand 
afterwards\footnote{
     If, in the end, this correction is deemed to be valid, 
     it may be worthwhile to revisit some of the pipeline files, 
     such as biases and darks, to see how they may have been 
     affected by imperfect CTE.  However, we will defer that 
     discussion until Section~\ref{s.SUMMARY}.}.

To test this shortcut, we first divided the {\tt flt} images 
by two to account for the fact that the {\tt flt} images are in 
electrons, while the corrections were developed for the {\tt \_raw} 
images, which represent DN with a gain of 2.  We then ran them 
through the same procedure as above, and multiplied the output 
by 2, to restore the scaling back to electrons.

The resulting images look extremely good; the trails appear to be 
removed nicely, just as well as in the raw images.  This makes sense 
in that the flat-field process should not introduce much more than 
a 5\% adjustment in the value of a pixel (the relative correction 
is typically less than 2\%).   While the CTE losses are not perfectly 
linear with pixel value, over a small range they are nearly linear, 
so it makes sense that operation on the {\tt \_flt} images would 
be nearly the same as on the {\tt \_raw} images.  In the remainder 
of this section, we make use of the {\tt \_flt}-stage routines to 
evaluate the efficacy of our correction in terms of absolute flux, 
astrometry, and shape.  As such, we will be reporting the results 
in electrons, not DN$_2$.

\subsection{Shallow-vs-Deep tests}
\label{ss.STARS_SHvDEEP}

It is well known that imperfect CTE affects short exposures with 
low background more than long exposures with higher background, 
so to test our correction we compared the short-exposure photometry 
and astrometry against the long-exposure photometry and astrometry.
Cycle-17 Program GO-11677 (PI-Richer, Co-I Anderson) will 
spend a total of 121 orbits staring with ACS at the outer 
calibration field in 47 Tuc, with the aim of detecting the 
end of the white-dwarf cooling sequence.  As of this writing, 
18 out of 24 visits have been successfully taken.  The 
observations in hand were taken in F606W and F814W, with 
exposure times typically between 1200 and 1300 s, depending 
on the visibility.  A single short exposure of 1 s, 10 s, 
or 100 s was taken at the beginning of each visit.  Each of 
these short exposures was followed immediately by a very 
deep exposure at the same pointing through the same filter, 
which will allow us to compare the CTE impact on the shorts 
and deeps.  The 1 s images did not have enough bright sources 
that could be seen unsaturated in the deep images to be useful, 
so we focus below on the 10 s and 100 s short exposures.  
For reference, the 10 s exposures have a background of about
2 e$^{-}$ and the background in the 100 s exposure was about 
15 e$^{-}$.

We took the three 10 s and the three 100 s short F606W exposures and 
their deep counterparts and corrected them for CTE according to the 
algorithm described above.  We ran the routine {\tt img2xym\_WFC.09x10}, 
which is described in Anderson \& King (2006), on the uncorrected and 
corrected {\tt \_flt} images This software program measures fluxes 
and positions for stars in individual exposures by fitting an empirical 
library PSF to the central 5$\times$5 pixels for each star.  The 
photometry it produces is equivalent to aperture photometry over the 
fitting box, with a correction for the fraction of the PSF that should 
have landed in the box, given the measured position of the star within 
the central pixel of the box.

\begin{figure}
\plotone{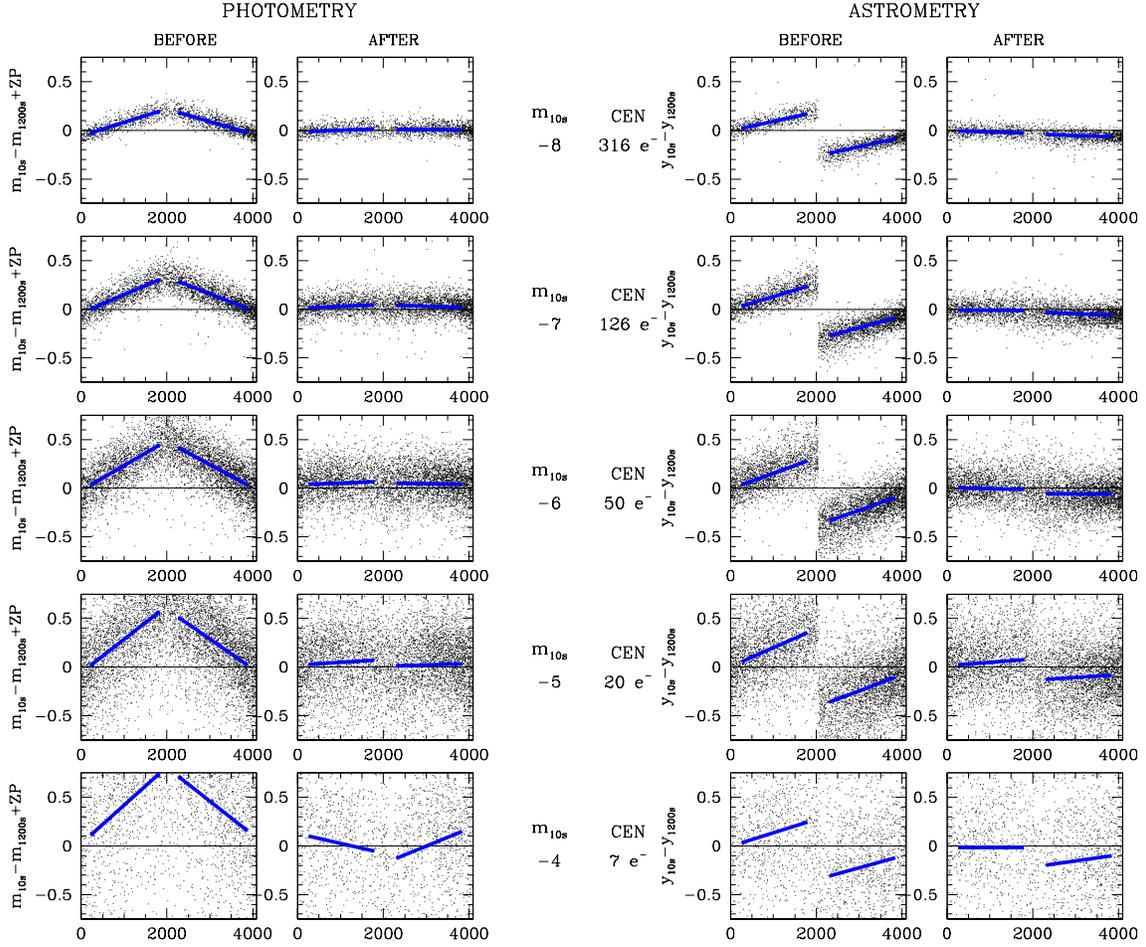}
\caption{Comparison of photometry and astrometry between 10 s and ~1200 s
         images for uncorrected and corrected images.  The photometric
         comparison is shown on the left, and the astrometric comparison
         on the right.  Photometry from the deep exposure has been 
         zero-pointed to correspond to the short exposure.  The residuals
         are plotted against the raw $y$ coordinate in a system that has
         the two chips abutted at $y=2048$.  The different rows of panels 
         show the residuals for stars with different short-exposure
         instrumental magnitudes.  Stars brighter than $-8.5$ in the 
         10 s exposures are saturated in the 1200 s exposures, so they
         could not be included.  The central column lists the instrumental 
         magnitude and the corresponding central-pixel intensity.  The 
         blue lines represent a linear fit for the trend against $y$.  
         \label{fig14}}
\end{figure}

\begin{figure}
\plotone{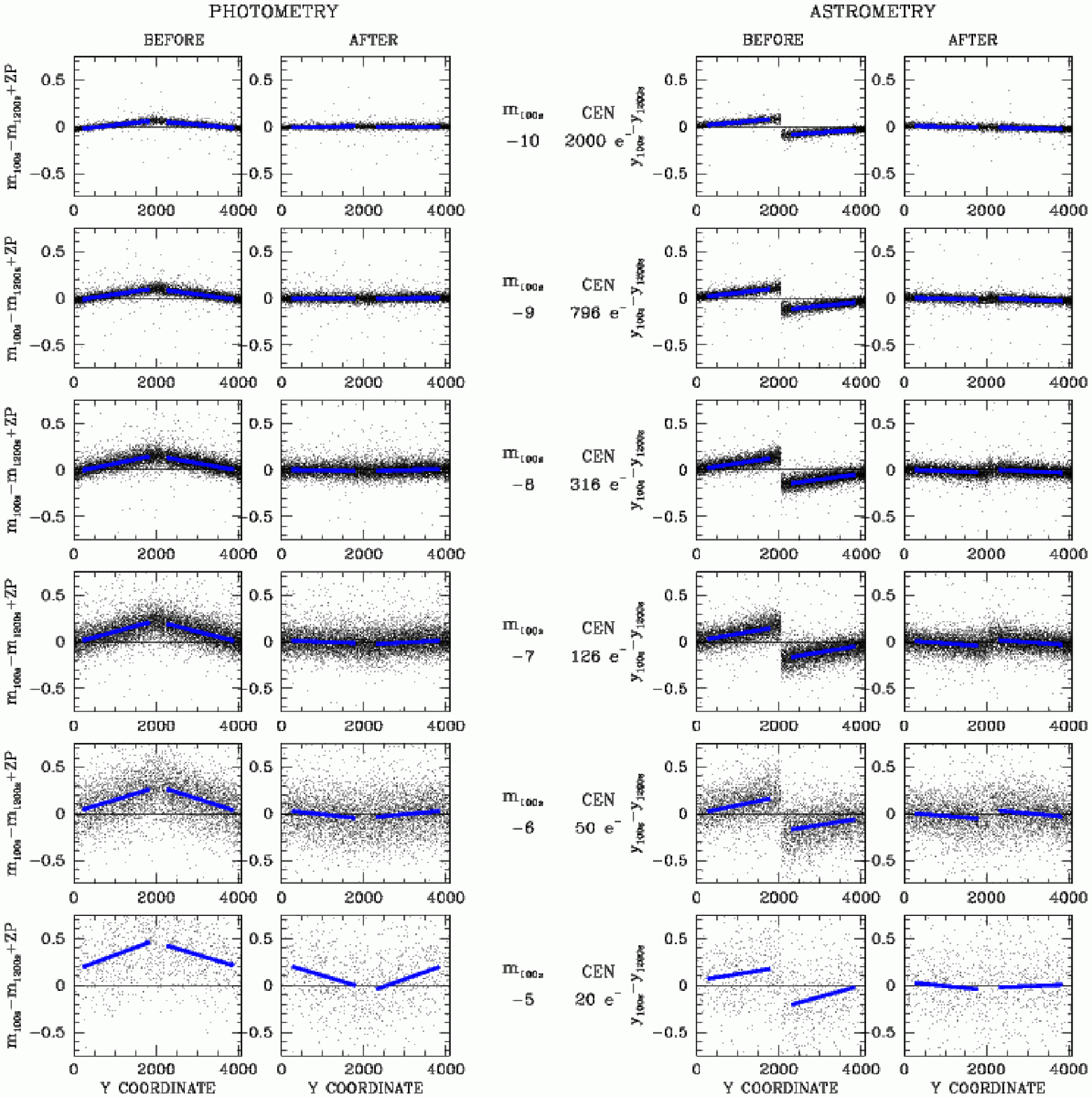}
\caption{Same as Figure~\ref{fig14}, but this time comparing the 100 s
         and ~1200 s exposures.  Stars brighter than $-10.5$ in the 100 s
         exposures are saturated in the deep exposures.
         \label{fig15}}
\end{figure}

By comparing the 10 s and 100 s images against their deep counterparts, 
we can examine directly how well our correction mitigates CTE in both 
astrometry and photometry.  Figure~\ref{fig14} shows the results from 
comparing the 10 s images against the $\sim$1200 s images for before 
and after CTE correction.  Figure~\ref{fig15} shows the same comparisons 
for the 100 s images.

The left column of panels of Figure~\ref{fig14} shows the photometric 
residuals for the 10s-vs-deep comparison.  The vertical scale covers 
$\pm$0.75 magnitudes in flux.  All the pre-correction comparisons show 
very significant CTE impact, with the familiar bow-shaped residuals 
indicating that stars at the center of the abutted two-chip system 
are farthest from the serial register and suffer more CTE loss in the 
short exposures than in the deep exposures.  The upper-right panel shows 
that the  brightest stars, with an instrumental magnitude\footnote{
    Instrumental magnitude is defined as
    $-2.5\ {\rm log}_{10}(FLUX)$.}
of $-8$, suffer a loss of about 0.2 magnitudes (20\%).  The lower-right
panel shows that the faintest stars at $m \sim -4$ suffer a loss of almost 
a 50\%.  Yet, in both cases the algorithm restores the photometry nicely, 
with an error generally less than 10\% of the correction itself.

The left panels of Figure~\ref{fig15} show the same for the 100 s 
exposures.  Stars in these medium-length exposures exhibit about a 
factor of two less CTE than stars of a similar brightness in the short 
exposures.

The panels on the right of both figures show the astrometric residuals 
along the readout direction ($y$).   Since the images were taken 
immediately after one another with the same guide stars, they should 
be almost perfectly co-registered.  The third column of plots exhibits 
the telltale sign of astrometric CTE (see Kozhurina-Platais et al.\ 2007):  
stars nearest the inter-chip gaps are farthest from the serial 
register and suffer CTE that shifts their position in the direction 
of the gaps.  The fourth column shows the corrected residuals, which 
largely remove the CTE signature.  Some of the intermediate-brightness 
objects in the 100 s exposures appear to be slightly over-corrected, but 
even there the correction is better than 75\%.

It is worth noting that we originally made this comparison using 
the corrected short exposures and the uncorrected deep exposures, 
thinking na\"ively that the deep exposures with their background 
of over 100 e$^{-}$ essentially made them immune to CTE losses.  
We found that without correcting the deep exposures as well, the 
short-vs-deep residuals indicated that the algorithm appeared to 
be over-correcting the short exposures.  However, when we applied 
the correction to both exposures, the correction was seen to be 
accurate and clearly relevant for both short and deep exposures,
as seen in Figures~\ref{fig14} and \ref{fig15}.

The fact that our correction restores photometry as well 
as astrometry is an indication that the flux we have associated 
with the trails is indeed almost {\it all} of the flux that is 
delayed by traps.  This was not an assumption of our algorithm; 
the aim of the algorithm was simply to restore the perceptible flux.
The fact that it appears to restore almost all the flux is 
satisfying.  This results is in contrast to what has been seen 
with WFPC2, where trail-based CTE corrections were deemed 
inadequate because they clearly did not account for all of the 
flux (see Figure~7 in Cawley et al 2001).  Of course the trail 
shapes and time constants can be very different for ACS/WFC and 
WFPC2, with their different array sizes, dopings, temperatures, 
and parallel-readout cadence.  It should not be surprising that 
they would have different trail profiles, and that WFPC2 with its 
faster readout (13 s vs 90 s) might have longer trails in a 
pixel sense.  We do note, though, that some investigations have 
shown that if the WFPC2 trails are modeled out to 90 pixels, the 
lost flux is in fact mostly accounted for (Biretta \& Platais 2005).

Laboratory FPR (first-pixel response) and EPER (extended pixel-edge 
response) tests have suggested that the observable trails do not 
contain all the flux.  However, such tests generally involve 
reading out only 20 or so virtual pixels, so they can measure 
the the flux only in the inner part of the trail.  Our trails 
have perceptible flux out to 60 pixels, so the conclusions from 
traditional EPER tests are not directly relevant.  Waczynski et 
al.\ (2001) made the same point, and found that when in their lab
experiments they extended their EPER test to 200 or 900 overscan 
pixels, they recovered the flux implied by the FPR tests.

\begin{figure}
\plotone{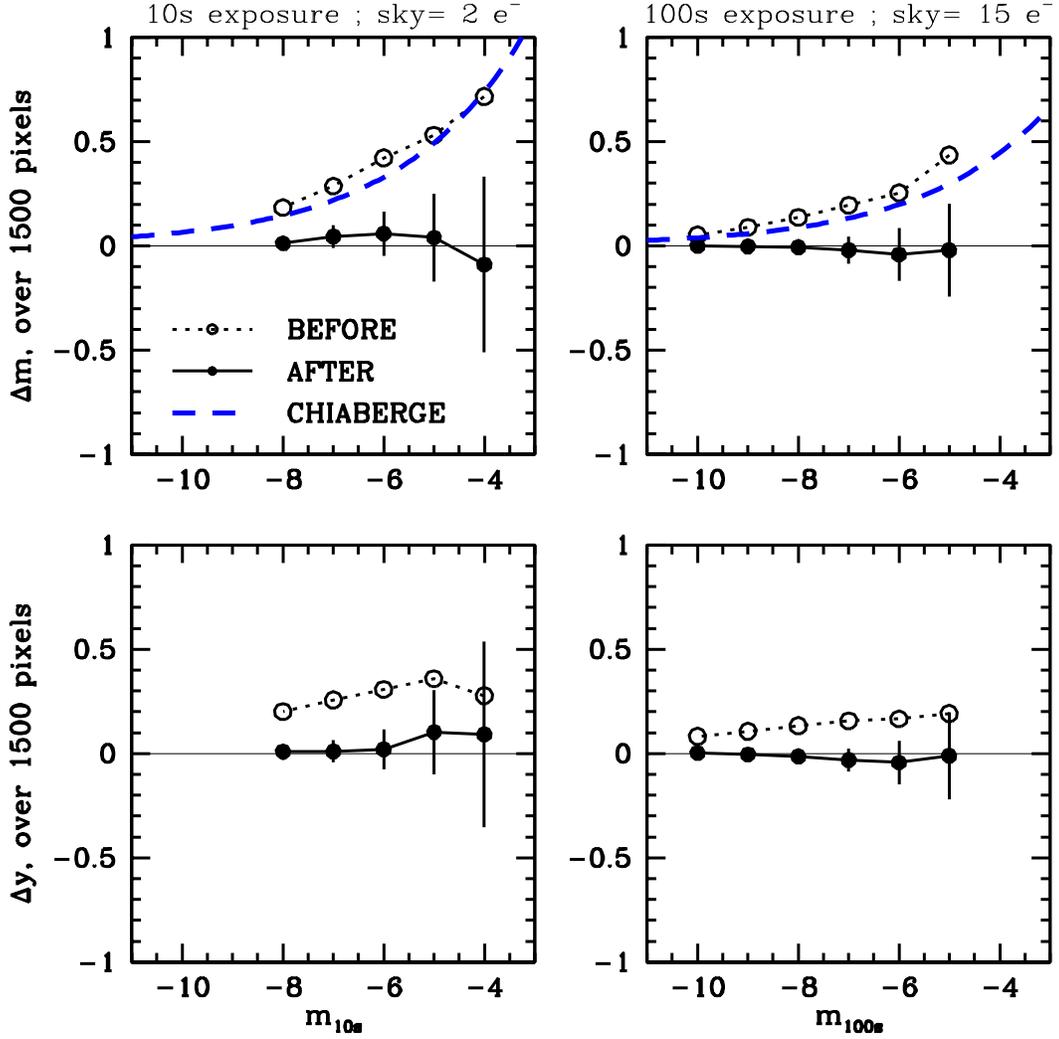}
\caption{The net photometric and astrometric corrections from our
         algorithm, in the upper and lower panels, respectively.
         The photometric corrections are compared against the 
         Chiaberge et al.\ (2009) correction for the appropriate date
         and background level.  The error bars indicate the spread
         in the residuals, not the error in the correction.
         \label{fig16}}
\end{figure}

In Figure~\ref{fig16} we compare the amplitude of our corrections
against those predicted by Chiaberge et al.\ (2009), plugging in for
the date and background of our exposures.  The aperture we used is 
effectively 2.8 pixels in radius, slightly smaller than the 
3-pixel-radius of the Chiaberge correction, but the results 
should not be particularly sensitive to this.   The upper panels 
show our net correction and the Chiaberge correction for the 10 s
and 100 s data set, as well as the trend seen in our correction.
The agreement is better than 20\% everywhere, and is additional 
encouragement that our correction restores essentially all of the 
flux.  The error bars drawn on our points reflect the photometric 
spread about the mean, not the error in the correction.  The bottom 
panels of the figure show the astrometric residuals, for which no 
Chiaberge-type empirical correction exists.
      
\subsection{The offset-image test}
\label{ss.OFFICIAL_COMP}

The ACS team regularly takes images to trend CTE losses.  One such 
program involves imaging the 47 Tuc calibration field with one 
pointing, then shifting the field vertically by one WFC chip so 
that the stars on the top chip in one exposure will be on the bottom 
chip in the other.  Such a strategy magnifies the impact of CTE, 
since stars farthest from the readout in one exposure will be closest 
in the other.

We have examined two such pairs of 30 s images taken in June 2009 
soon after SM4:  {\tt ja9702sxq}, {\tt ja9702szq}, {\tt ja9702t5q} 
and {\tt ja9702t7q}.  The images have a background of about 4 e$^{-}$. 
We measured the stars in all the exposures (corrected and uncorrected)
using the same photometry routine as above, and compared the photometry 
for stars found in all four exposures as a function of where the star 
was found in the bottom-chip exposure.

\begin{figure}
\plotone{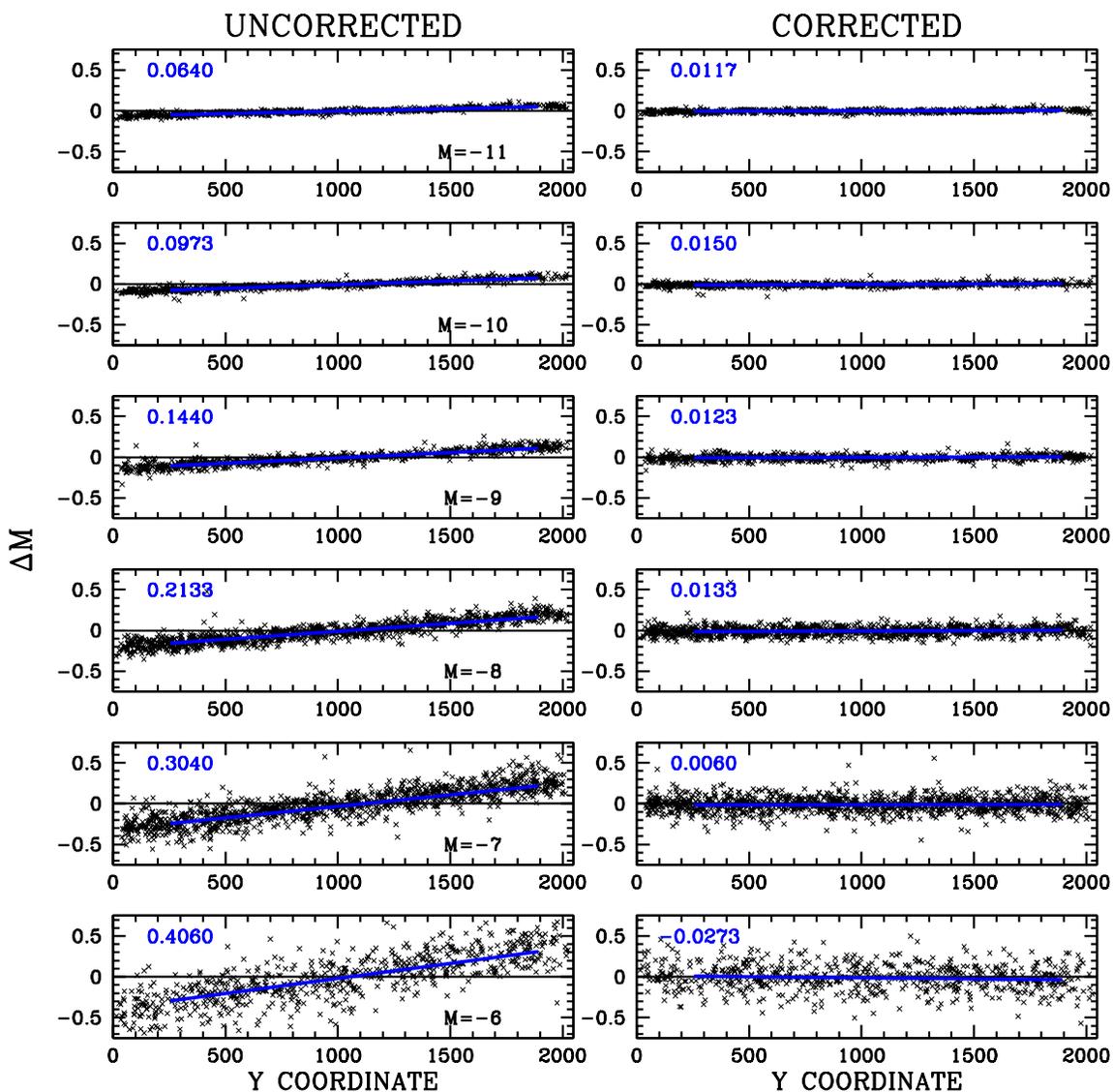}
\caption{Comparison of the measured brightnesses of stars in images 
         that have been shifted by a full chip to enhance CTE impact. 
         Uncorrected images are shown on the left, and corrected images
         on the right.  The panels from top to bottom shows stars with:
         25000, 10000, 4000, 1500, 630, and 250 $e^{-}$ total flux.
         The blue lines indicate the linear trend, and this slope 
         (divided by two, to account for the two-image comparison)
         is indicated in the upper left of each panel.
         \label{fig17}}
\end{figure}

The results of this test are shown in Figure~\ref{fig17}.  The size
of the effect here should essentially be twice that seen in a single
exposure.  We see that before corrections, the relatively bright stars 
that are farthest from the amplifier in one exposure and closest to 
it in the other differ by 12.8\% (i.e., $2 \times 0.064$).  The fainter 
stars differ by more than 50\%.  After the corrections, there is very 
little discernible trend.

\subsection{More absolute tests}
\label{ss.ABS_TESTS}

We have also looked into more ``absolute'' tests, and compared
positions measured in individual exposures of the 47 Tuc calibration
field against photometry and astrometry in a reference catalog that 
has been constructed from many pre-SM4 exposures taken at a variety 
of orientations and offsets.  The photometric results were very 
encouraging, but not perfect, perhaps indicating that the reference 
catalog has been compromised more than anticipated by CTE.  In an 
astrometric sense, the stars in the reference frame are all moving 
with internal motions of about 0.5 mas/yr relative to each other 
due to the cluster's internal motion (McLaughlin et al.\ 2006), 
making the astrometric handle fuzzy.

Over the next few months, as this routine is evaluated for use
before SM4 and before the temperature change, we will apply it
to the archive exposures and construct a better catalog, which can
provide a photometric ``truth'' against which observations can
be compared directly.  Also, the GO-11677 data set will be ideal 
for examining the CTE correction and its efficacy for faint 
sources on moderate backgrounds.

\subsection{Comparing shape}
\label{ss.STARS_SHAPE}

The fact that the astrometric bias is generally removed by the CTE 
correction encourages us to think that the correction might preserve 
shape as well.  This aspect of the correction will be important for 
studies of lensing and galaxy morphology.  We study this more 
quantitatively by examining the profiles of stars in the corrected 
and uncorrected deep and shallow images.

We began by mapping the star catalog from Anderson (2007) 
into the {\tt \_flt} frames of {\tt ja9bw2ykq} and {\tt ja9bw2ylq}, 
which had exposure times of 30 s and 339 s and backgrounds of 3 and
35 $e^{-}$, respectively, and were taken immediately one after the
other with no dither offset.  We analyze only the stars between 
$y = 1500$ and $y = 2000$ of the bottom chip in an effort to 
maximize our sensitivity to CTE-related issues.  

The catalog was constructed in 2005.  Using the exact 2005 positions 
in 2009 images would introduce proper-motion-displacement errors to 
our positions, so we used the catalog only to identify {\it bone fide\,} 
stars.  We used the library PSFs described in 
Anderson \& King (2006) 
to determine an 
improved position and flux for each star in the deep corrected image.
We then extracted a 5$\times$5 raster centered on each star from each
of the four images (the corrected and uncorrected deep and shallow 
exposures).  We also removed from each raster a sky as measured 
from the annulus between 8 and 12 pixels.  Finally, we examined 
these pixels to see how imperfect CTE had modified the stellar 
profiles, and to determine whether our corrections had restored them.

\begin{figure}
\plotone{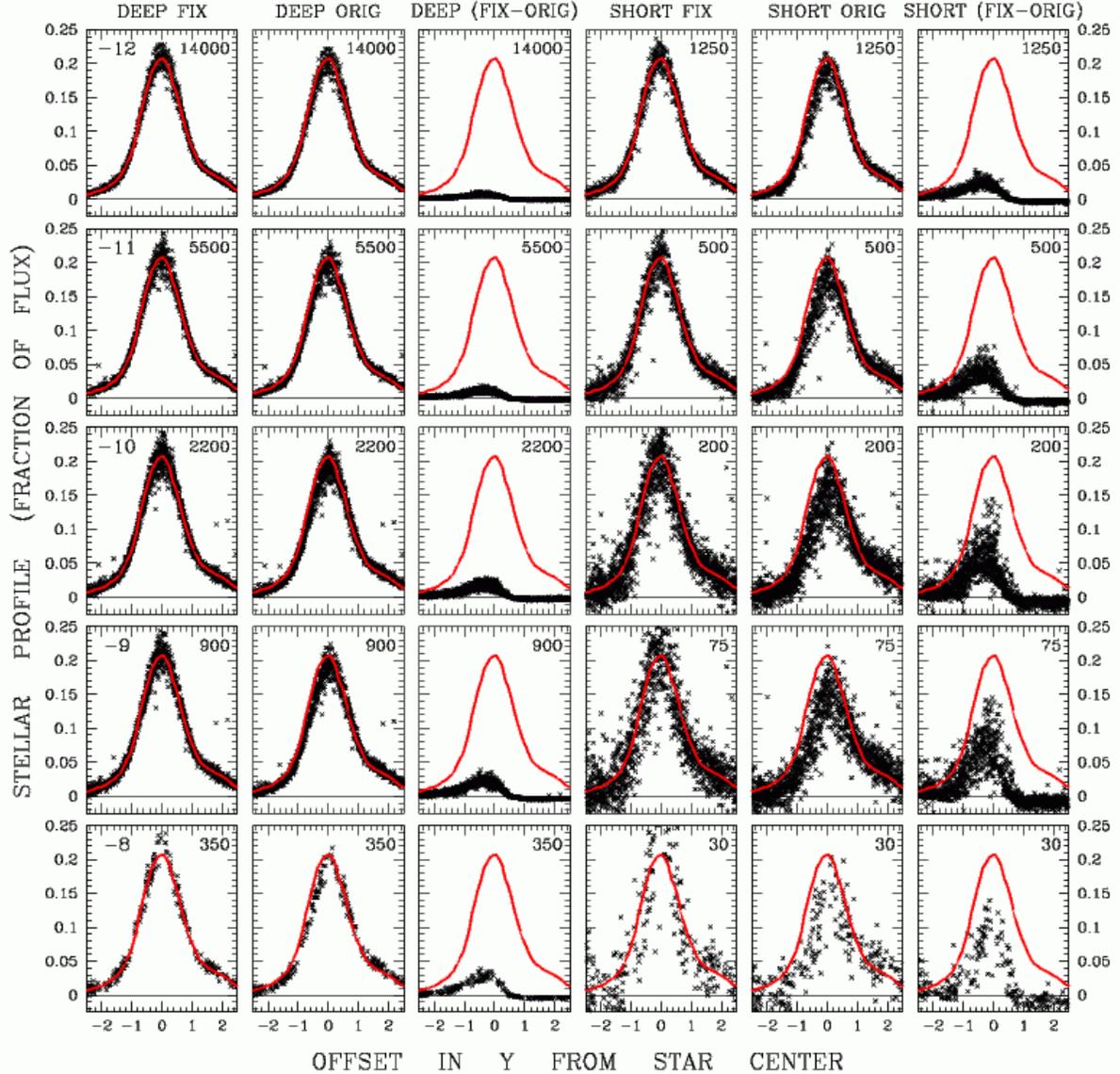}
\caption{The profiles of stars in corrected and uncorrected deep (339 s) 
         and short (30 s) images of the 47 Tuc calibration field.  The
         first and second columns show the corrected and uncorrected 
         deep exposures.  The third column shows the difference, which
         corresponds directly to the CTE correction.  The next three
         columns show the same, but for the shallow exposure.  The
         brightest stars are shown in the top row, and the faintest 
         in the bottom row, with the deep-image instrumental magnitude 
         listed in the upper left of the left plots.  The flux in the 
         central pixel of a typical star is listed (in $e^-$) in the 
         upper right of each panel.  The PSF profile (red solid line) 
         drawn in each plot comes from tracing the distribution in the 
         upper-left panel, and reproducing the same curve in the other 
         panels.  The units of the vertical axis are fraction of 
         flux in each pixel, relative to the total within a 
         0.5\arcsec aperture.
         The horizontal axis in each goes from $-2.5$ to $2.5$ in WFC 
         pixels (along $y$).
         \label{fig18}}
\end{figure}

Anderson \& King (2000) showed how one can directly examine the 
PSF profiles of stars if we have an estimate of their fluxes and 
positions (see their Figure 6 and Eqn.\ 7).  From the deep corrected
exposures, we have estimates of the stars' positions and fluxes, so
we can examine their profiles.  In Figure~\ref{fig18} we plot 
the vertical profile of the central part of the PSF for stars of 
different brightness in the different images.  Specifically, we 
plot $\hat{\psi}(\Delta x,\Delta y)$ against $\Delta y$ for 
$|\Delta x| <0.5$.

The PSF as formulated here simply maps the fraction of light that 
falls in each pixel as a function of where the pixel is relative 
to the center of the star.  For example, $\psi(0,0)$ represents 
the fraction of light that would fall in a pixel when the star 
is centered on that pixel.  For the WFC PSF in F606W, between 
19\% and 25\% of the light in the star falls in the central pixel 
(Anderson \& King 2006).

All stars bright and faint should have the same PSF (leaving aside
issues of minor position-dependent variations).  The upper-left panel 
shows the profile for the brightest stars in the corrected deep image.  
This should be representative of the true PSF profile, so we trace 
it with a red line and plot it for reference in all the other panels.
Going down, we show that the stars from instrumental magnitude $-12$
to $-8$ in the corrected image all have the same profile, in that the
points (each of which represents one pixel in one star image) all
trace out the same curve.  The brightest stars have 14000 $e^{-}$ in
their central pixels, and the faintest have 350 $e^{-}$.  All the 
panels in each row show the exact same stars.

The next column over shows the profiles from the untouched deep 
exposure.  This by and large traces the same profile, but as the 
source gets fainter, it is clear that the leading edge (on the left) 
suffers some CTE losses.  The third panel shows the difference 
between the two profiles, corresponding directly to the adjustment 
that the CTE algorithm has made to the original pixel values.

The fourth panel shows the profiles from the stars in the corrected
short exposures.  The brightest stars here have 1250 $e^{-}$ in their
central pixels and the faintest have only about 30 $e^{-}$.  It is
clear that the corrected pixels follow the profiles extremely 
closely---even for the faintest stars.  Our correction not only 
restores the flux but also the shape of astronomical sources.  

The second to last column shows results for the untouched short 
exposure.  It is easy to see here how imperfect CTE has shifted and 
modified the stellar profile:  a larger and larger fraction of 
electrons from the leading edge of the star are lost as the star 
gets fainter.  By contrast, there is a clear excess of flux in the 
CTE trail on the right.

The adjustment made by our algorithm is shown in the rightmost column.
The faintest stars require an adjustment that corresponds to nearly 
50\% of the original flux.

Our CTE restoration algorithm was constructed entirely from the 
profiles of warm pixels in dark exposures, with no reference
whatsoever to stars or their profiles.  The fact that this algorithm 
restores the pixels of a stellar profile to their rightful values 
is a clear independent demonstration that the algorithm not only 
corrects stars' fluxes, as we saw in \S\ref{ss.OFFICIAL_COMP}, but 
puts the flux in the right place as well.  This gives us every reason
to think that the same should be true for resolved galaxies, which 
are even less sharp than PSFs.



\section{POSSIBLE IMPROVEMENTS TO THE ALGORITHM}
\label{s.IMPR}

The preceding sections have shown that our CTE-correction algorithm 
clearly works quite well.  The process of restoring flux to the 
bright sources naturally lowers and improves the uniformity of the 
background as well.  This can be seen from a simple inspection of 
the images.  However, there are still some remaining issues, such as:  
read-noise amplification, the speed of the algorithm, and CTE issues 
in the serial direction.  We will address these below.

\subsection{Readnoise Amplification}
\label{ss.IMPR_RNMIT}

The imperfect charge-transfer process tends to blur out the images.
What is originally a delta-function warm pixel gets read out as a 
less intense delta function with a tail.  In general, this blurring 
will be mild.  If we have a single bright pixel with more than 15 DN$_2$ 
on a negligible background, most of the electrons will stay with the 
original pixel packet, and we can hope to re-associate the trail-electrons 
with their original packet.  

Nevertheless, the reconstruction process still constitutes a 
deconvolution, however mild, and as such it should be expected that 
a reconstructed image will end up being sharper than the image that 
was read out at the amplifier.  Therefore, to the extent that the observed 
image has any variation from pixel to pixel, the reconstructed image 
will necessarily have more.  So long as this ends up restoring true lost 
sharpness to stellar profiles or object morphology, this sharpening 
is appropriate.  But there is one component of the observed image that 
did not participate in the charge-transfer process:  the read-out 
noise (RON).

The RON in ACS was 5.5 $e^{-}$ before SM4 and has been 3.9 to 4.7 $e^{-}$ 
since SM4 (Maybhate et al.\ 2010).  This noise gets added to the 
image pixels as they are read out at the amplifier.   It was not 
present during the pixel transfer, so if we invert the readout process 
on an image with RON, we will arrive at an image with a scaled-up 
version of the readnoise, since this is the only way to arrive at 
the observed image after the blurring readout process.  To estimate 
this amplification factor, we generated an image with a flat background 
of 5 DN$_2$ and a readnoise of 2.25 DN$_2$, subjected this image to 
the readout-inversion process and ended up with an image that had a 
pixel-to-pixel noise of 3.06 DN$_2$, thus increasing the readnoise 
by about 40\%.  This is not desirable, but it would seem unavoidable, 
as we would like to neither remove the unavoidable noise arbitrarily, 
nor amplify it.  

Although it is never possible to know precisely what pixel-to-pixel 
variation in an image is due to readnoise, Poisson noise, or actual
structure in the scene, it is possible to take a conservative approach:
we will attribute as much of the pixel-to-pixel variation as possible to 
readnoise and operate our procedure on the rest.  To do this,
we will separate the observed image $I_{\rm OBS}$ into two components 
$I_{\rm OBS} = I_{\rm SM} + I_{\rm RON}$.  The first component is a 
slightly smoother version of the observed image; the second component is 
an image that is consistent with being pure readnoise.  As such, 
it has no large-scale structure with only high-frequency 
pixel-to-pixel variations with an amplitude of ($< 5$ DN$_2$).

We effected this separation by first taking the original image and 
smoothing it with a 3-pixel boxcar in $y$.  We subtract this smooth 
image from the original to obtain the high-frequency component.  Any 
pixels in this component less than 5 DN$_2$ are likely to be indicative
of read-noise.  High-frequency-component pixels greater than 
10 DN$_2$ are likely due to something other than RON
(e.g., Poisson noise, real high-frequency structure in the 
astronomical scene, a WP, or a CR).  So, to get the best estimate
of the pure RON component, we map this residual $\delta(i,j)$ into 
$\Delta(i,j)$, something more reflective of something that is pure 
readnoise:
\begin{equation}
\Delta = \left\{
       \begin{array}{rcl}   0: & {\rm if} &       \delta < -10 \\ 
                   -10-\delta: & {\rm if} & -10 < \delta <  -5 \\ 
                       \delta: & {\rm if} &  -5 < \delta <  +5 \\
                    10-\delta: & {\rm if} &  +5 < \delta < +10 \\ 
                            0: & {\rm if} &       \delta > +10 
       \end{array}. 
\right.
\end{equation}
If $|\delta| < 5$, then the high-frequency component for that pixel 
is consistent with being entirely readnoise, so $\Delta$ = $\delta$.  
If $|\delta| > 10$, then it is determined to be something other than 
readnoise and $\Delta$ is set to zero.  We taper $\Delta$ between 
these two extremes.  Finally, we associate the readnoise image 
with $\Delta$, $I_{\rm RON} = \Delta$, and the smoothed image is then 
simply $I_{\rm SM} = I_{\rm OBS} - \Delta$.

We then execute the pixel-restoration procedure on the smoothed image,
$I_{\rm SM}$, to get the corrected-smooth image, $I^{\prime}_{\rm SM}$.  
Once this is constructed, we add back in the readnoise component to 
get the best estimate of the corrected image, $I^{\prime}_{\rm ORIG}$.
In this way, we aim to operate the restoration algorithm only on 
the part of the image that is clearly not readnoise, while at the same 
time, modifying the ultimate noise properties as little as possible.  

To test this algorithm, we took a short 30 s exposure of the 47 Tuc 
calibration field ({\tt ja9bw2ykq}), which has a background of 3 $e^{-}$ 
and reconstructed it two ways:  (1) we applied the reconstruction routine 
on the original image, and (2) we applied the reconstruction routine on 
a smoothed version of the original image, then added in the un-smooth part:  
$I^{\prime}_{\rm ORIG} = I^{\prime}_{\rm SM} + I_{\rm RON}$.

\begin{figure}
\plotone{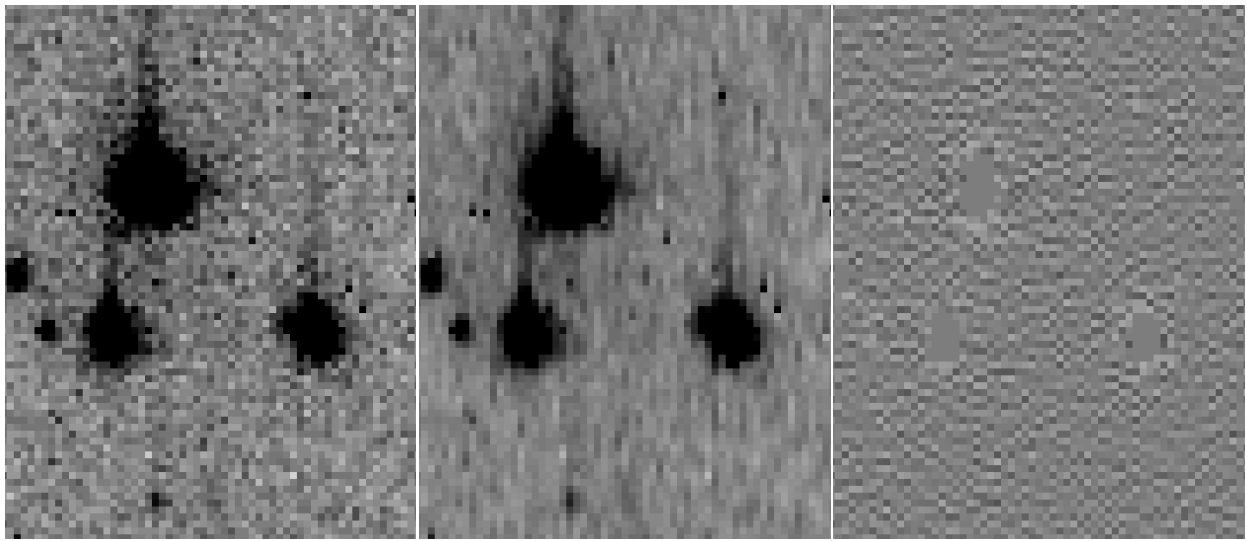}
\caption{(Left) Portion of image {\tt ja9bw2ykq\_raw} 
                (background 1.5 DN$_2$) centered on coordinate 
                (1215,1609) in the bottom chip.
         (Middle) The smoothed component of this image.
         (Right)  The component of the image attributed to 
                  readnoise.
         \label{fig19}}
\end{figure}

Figure~\ref{fig19} shows a portion of the observed {\tt \_flt}
image, and its smooth and unsmooth components.  The unsmooth image 
has a total range of $\pm$5, and is zero in regions of real structure 
(near the locations of stars).  

\begin{figure}
\plotone{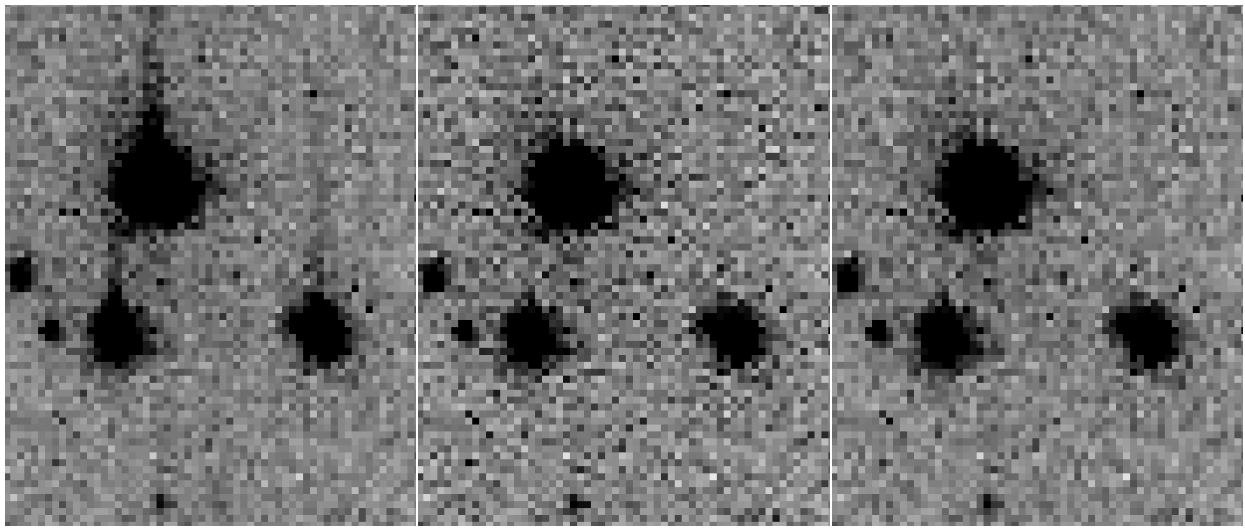}
\caption{(Left)   The original image.  
         (Middle) The direct reconstruction of the original.
         (Right)  The reconstruction based on the smoothed image.
         \label{fig20}}
\end{figure}

Figure~\ref{fig20} shows the two reconstructions.  The original
image is shown on the left, with the direct reconstruction in the middle
and the smooth-based reconstruction on the right.  It is clear that the
background is much noisier in the middle exposure than in the original
or in the reconstruction on the right.

\begin{figure}
\plotone{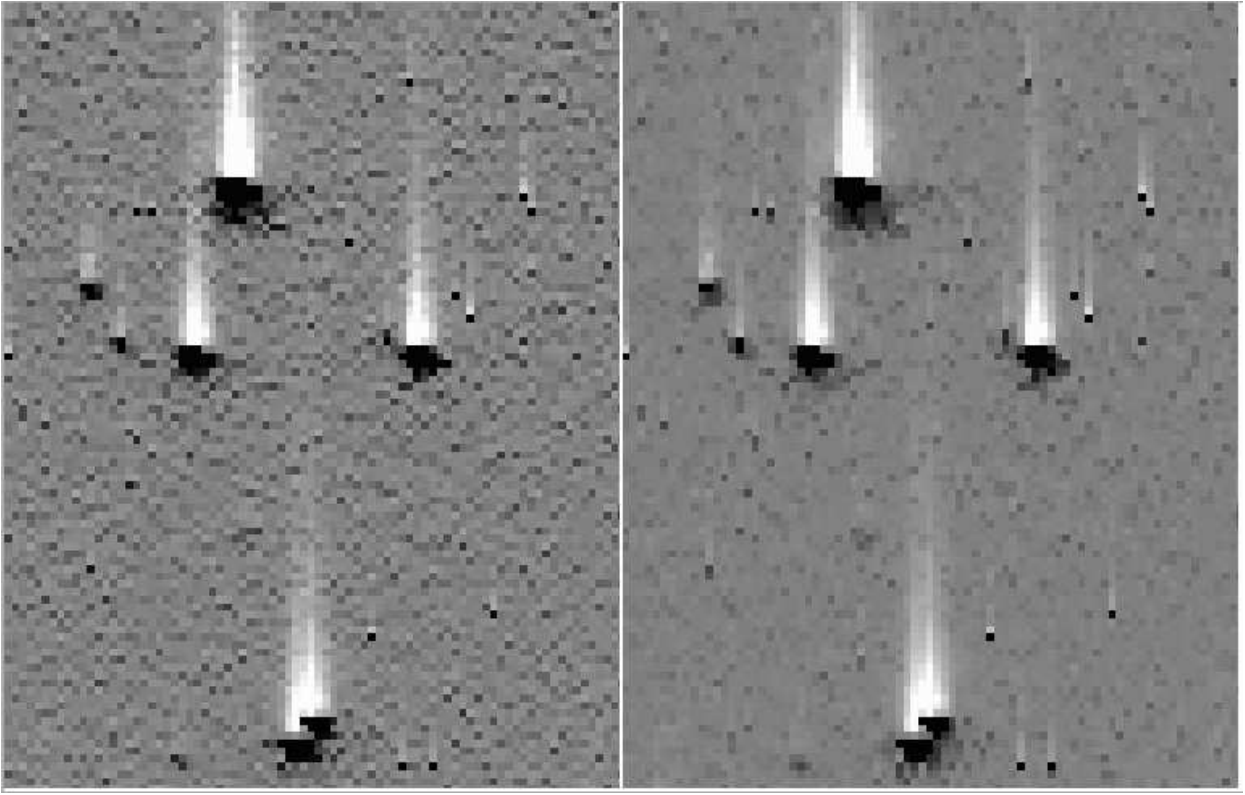}
\caption{The CTE adjustment made to the original image for the
         direct reconstruction (left) and for the smooth-image-based
         reconstruction (right).  
         \label{fig21}}
\end{figure}

\begin{figure}
\plotone{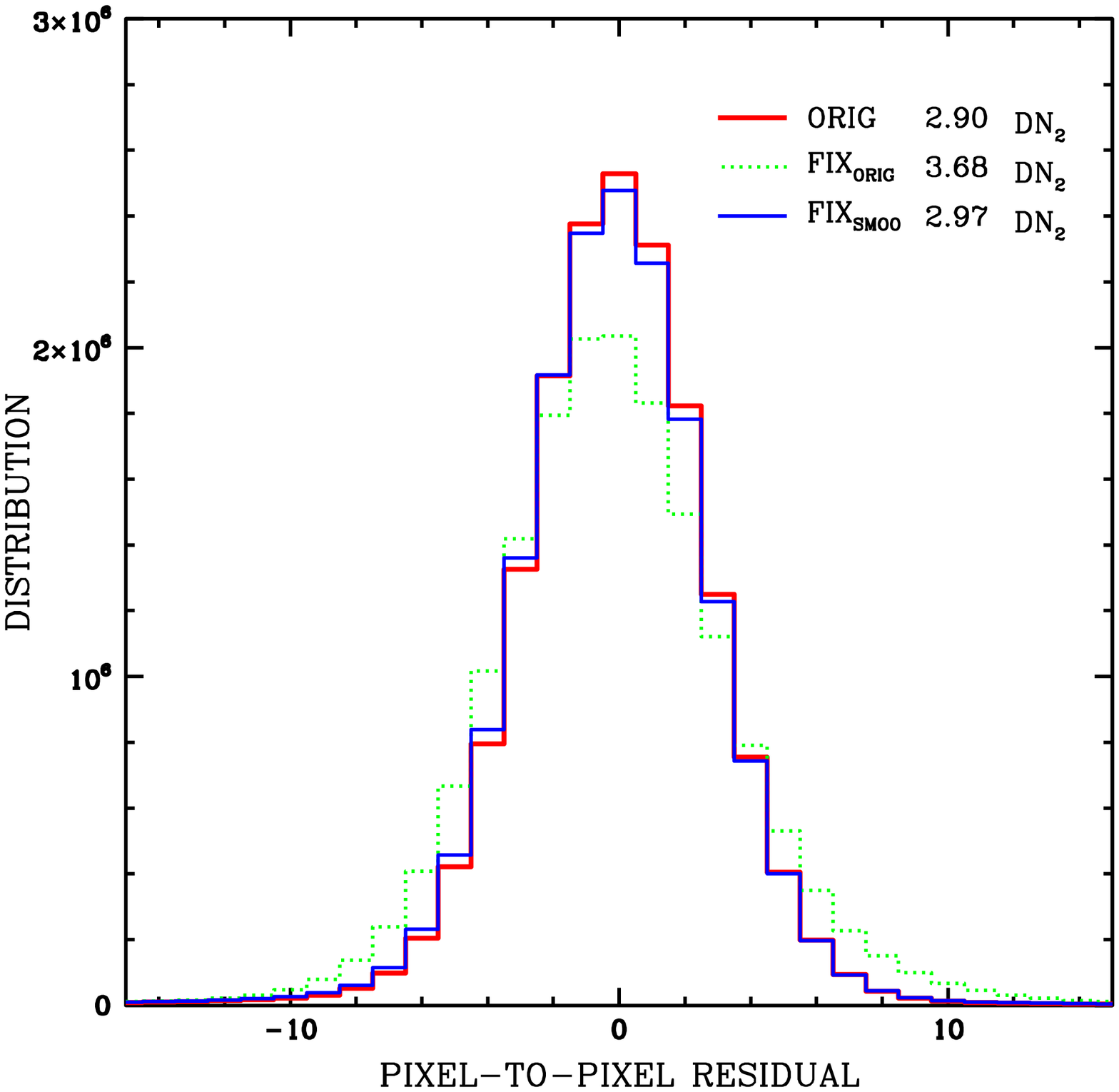}
\caption{Histograms showing the pixel-to-pixel variation in the 
         original image (heavy red), the direct-restored image 
         (dotted green), and the smooth-image-based restoration (blue).
         The residual is computed for each pixel and is simply the
         difference between the pixel and the average of its eight 
         surrounding neighbors.
         \label{fig22}}
\end{figure}

Figure~\ref{fig21} shows the actual difference between the original
image and the two restored images.  It is clear that the two have similar
effects on the stars, WPs, and CRs, but the direct reconstruction clearly
adjusts many more pixels in the background.  Figure~\ref{fig22} shows
this in histogram form.  The restoration acting on the original image 
increases the pixel-to-pixel variation by about 30\%, while the
restoration based the smoothed image increases it only by 3\%.  

This multi-step restoration will make a big difference for images 
where the background is low, but will make only a small difference
in images where the noise is dominated by Poisson photon statistics, 
since this source of noise {\it has\,} participated in the CTE-impacted
charge transfer.  The algorithm presented here may not be the only 
one way to address the readnoise-amplification issue, however it does
demonstrate that it should be possible to restore CTE-trails without 
introducing additional unpleasant artifacts.

\subsection{Improvements in Computational Speed}
\label{ss.IMPR_FASTER}

The prescription given for the algorithm in \S\ref{ss.MODEL_ALGO} 
involves dealing with fractional charge traps at every marginal unit 
of DN$_2$.  This means that we need to monitor the state of each of 
up to 40,000 traps, as the flux is transferred from one pixel to another.  
When operating on typical images with backgrounds of about 25 DN$_2$, 
the routine takes about an hour on a 2.4 GHz machine to complete its 
five reconstruction iterations for all the 4096 columns of both chips.  
This is not prohibitive, but if it could be sped up without any
loss of accuracy, it would be a tangible benefit.

The function $\phi_q$ tells us how much charge the trap at each 
DN$_2$ level ($q$) can affect.  Adding $\phi_q$ up from $q=1$ DN$_2$ 
to $q=$50,000 DN$_2$, we find that a full-well pixel is subject to
about 1000 DN$_2$ worth of traps as it is transferred from the top 
to the bottom of the detector (see Fig~\ref{fig06}).  On average, 
then, the trap at each individual level of $q$ affects only 
0.02 DN$_2$ of charge.  Since the readnoise is about 2.5 DN$_2$, 
it does not make sense to track the charge traps at this resolution.
We can instead afford to parametrize the traps 
more intelligently.

We modified our routine to step through the traps not by every DN$_2$ 
in pixel value, but rather by every block of marginal charge that amounts to 
one DN$_2$ of traps.  As a result we end up with a total of about 
1000 traps, one at 1 DN$_2$, one at 2DN$_2$, one at 4DN$_2$, then at 
5, 6, 8 DN$_2$, 10, 11, 13, 14, 15, 17 DN$_2$, etc.  The trap density 
is higher at lower charge levels, so at higher levels we skip for 
instance from 1000 to 1012.  It is much quicker to keep track of only 
1000 traps than 50,000.  The difference in the model output 
is negligible, but the difference in execution time is enormous.  
We found that after this small change to the code, it ran about 
6 times faster, doing 5 iterations on an entire 4096$\times$4096 WFC 
image in about 10 minutes --- a considerable improvement.

\subsection{CTE in the Serial Direction}
\label{ss.IMPR_XCTE}

\begin{figure}
\plotone{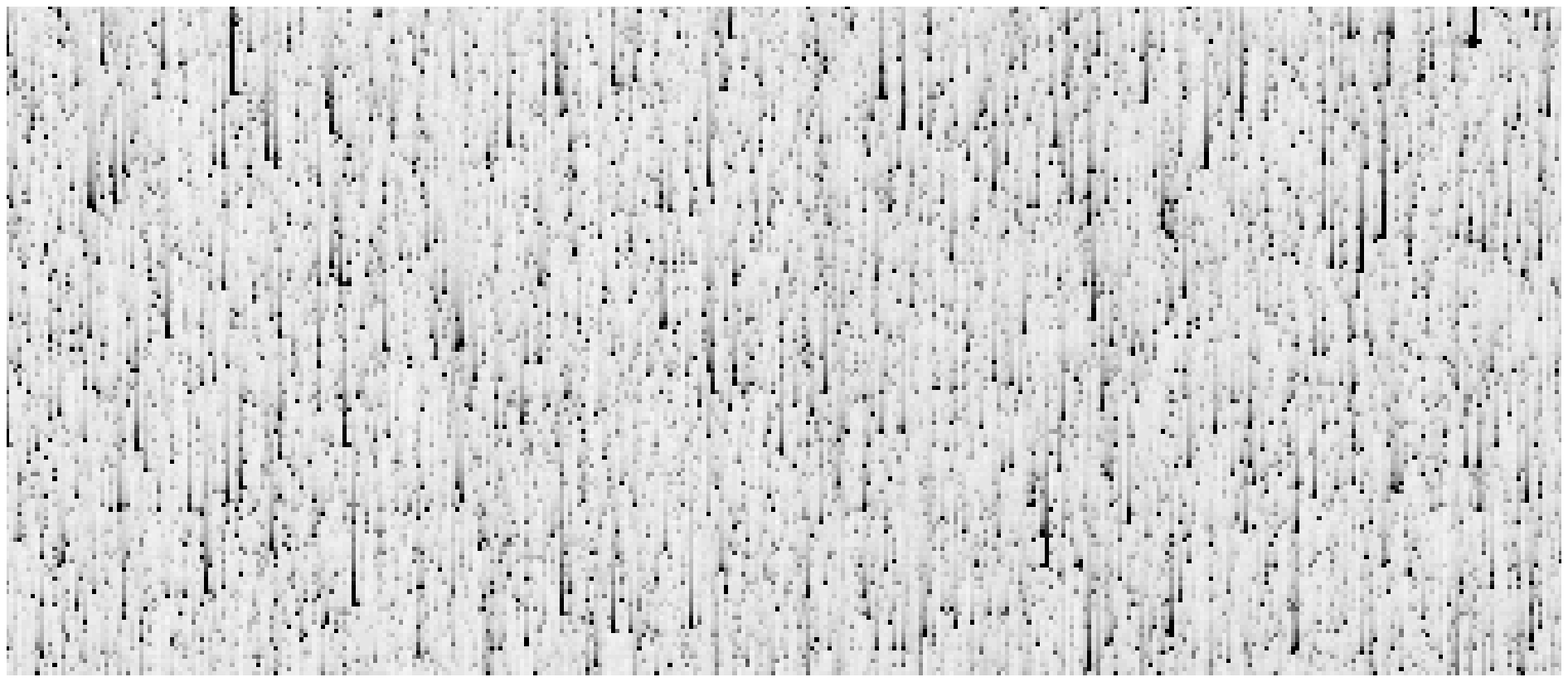}
\caption{The region of the dark stack centered on pixel (1860,2172).
         The bright WPs show a clear $x$-hook towards the center 
         of the image and away from the amplifier.  Pixels are 
         shifted leftward in the left half of the image, and rightward 
         in the right half of the image.
         \label{fig23}}
\end{figure}

The possibility of imperfect CTE along the serial register (in the $x$ 
direction) has been looked for in ACS images, but it has been found 
to be so low as to be unquantifiable (Riess \& Mack ISR-2004-06).  A 
cursory look at the dark stacks that we generated in \S~\ref{ss.DARK_ANAL} 
shows a curious $x$-hook in the direction away from the readout 
amplifier for the brighter warm pixels.  Figure~\ref{fig23} shows 
a close-up of the region about $i=2072$, which is the boundary where
the serial readout direction changes from right-to-left to left-to-right 
in the full 4144$\times$4136 images (See Figure~\ref{fig02}).  Below, 
we apply the same methodology to examine X-CTE that we used above for Y-CTE.  

We return to the dark stack image from \S\ref{ss.DARK_ANAL}.  For each 
of the significant warm pixels (i.e., those with peak-map hits of 125 or 
greater), we recorded the array of pixels from 5 pixels to its 
left to 5 pixels to its right.  The schematic in Figure~\ref{fig04}
shows which pixels were used to measure the serial trails.  We 
constructed the empirical trails by subtracting the downstream 
pixels from the corresponding upstream pixels to remove the 
background.  The excess in the first pixel in the trail can be seen 
by looking at $XU1-XD1$, etc.

\begin{figure}
\plotone{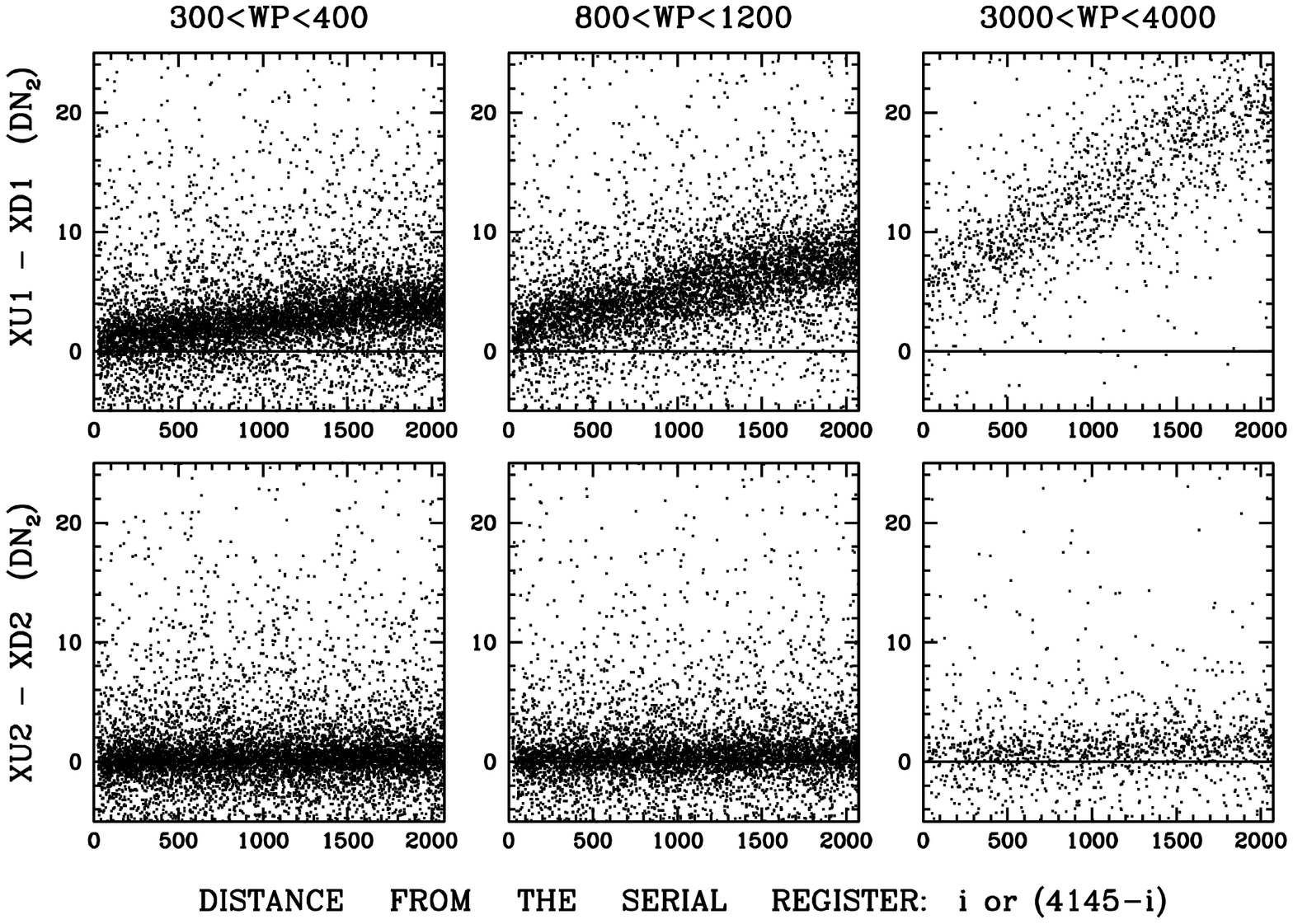}
\caption{(Top)    For three different intensities of warm pixel, we 
                  show the number of DN$_2$ in the first pixel in 
                  the serial trail as a function of the number serial 
                  shifts to the readout amplifier.
         (Bottom) We show the same for the second pixel in the serial
                  trail.
         \label{fig24}}
\end{figure}

Figure~\ref{fig24} shows the first and second pixels in the trail as 
a function of distance from the serial register.  The three panels 
from left to right show the trend for three different warm-pixel 
intensities.  Clearly, the brighter the warm pixel, the brighter 
its trail.  It is clear that there is a simple linear correlation 
between the first-pixel intensity and the distance from the serial
register---a clear telltale sign of CTE loss.  Surprisingly, the 
intercept in the relation is not at $i=0$ but somewhere around 
$i=-500$, which is likely due to an effect caused by the new ASICs 
called the ``bias-shift'' (see Golimowski et al.\ 2010 for a discussion.)

\begin{figure}
\plotone{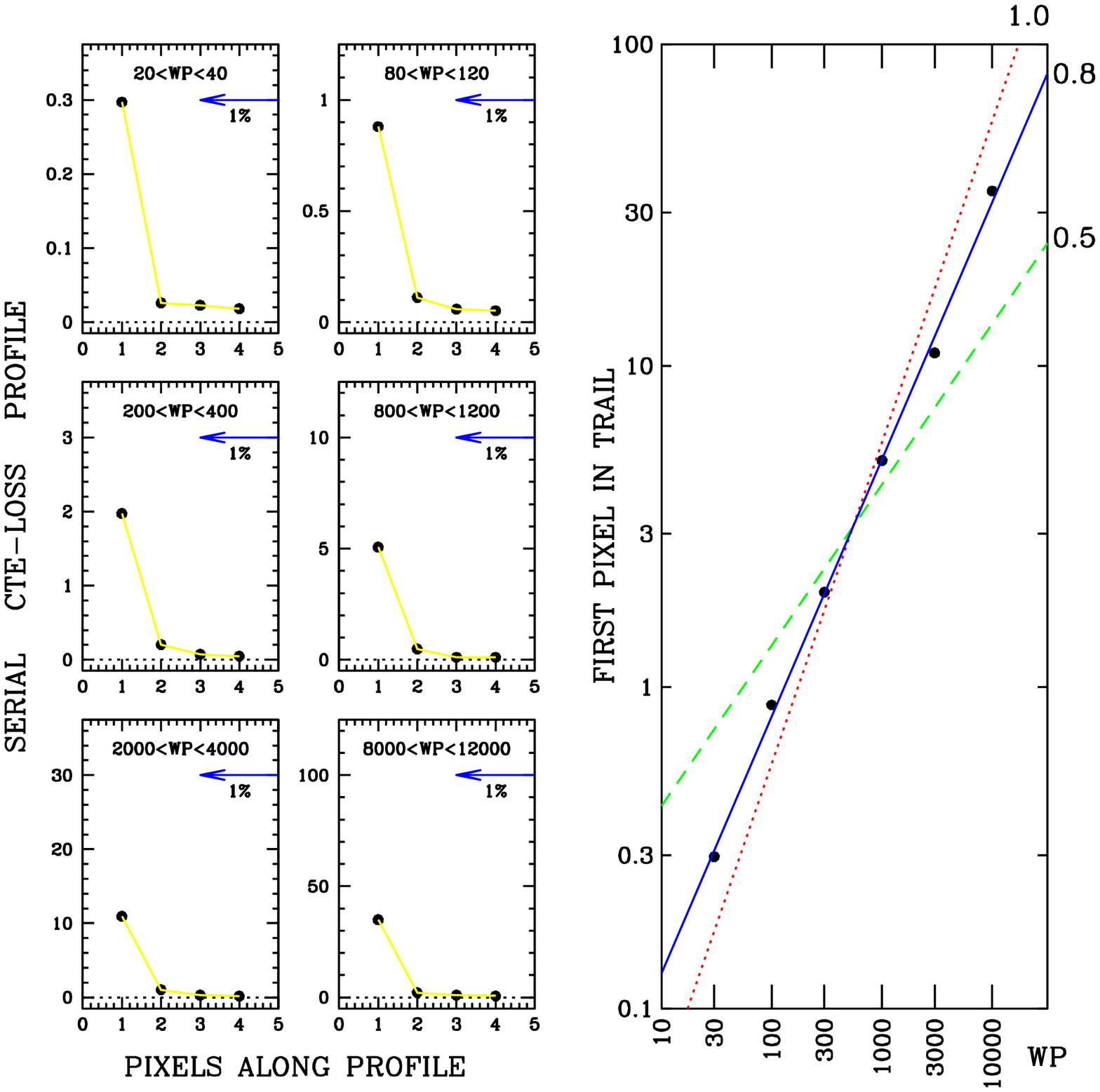}
\caption{The panels on the left show the first four pixels in the 
         serial-CTE trails for six different warm-pixel intensity bins, 
         labeled at the top of each panel.  The arrow in each plot 
         shows the 1\% level relative to the WP intensity:  the 
         first pixel intensity goes from 1\% at $\sim$30 DN$_2$ to less
         than 0.4 \% at $\sim$10,000 DN$_2$.  The right panel shows the 
         intensity of the first pixel in the trail as a function 
         of the warm pixel intensity.  The slopes for the power laws 
         drawn in are indicated on the upper right.
         \label{fig25}}
\end{figure}

Figure~\ref{fig25} shows the profile of the X-CTE trails for six different
bins of WP intensity.  The first-pixel intensity goes from about 1\% of 
the WP at the faint end, to about 0.4 \% at the bright end.  There is 
very little noticeable flux in the second and subsequent pixels, so the
serial CTE trails are quite different in nature from the parallel trails.
This makes sense, as the dwell-time for charge packets in the parallel 
shifts is more than 2000 times longer that for the serial shifts 
(22 $\mu$s).  This result is also consistent with expectation from 
theory and laboratory experiments, which show that the trap-capture 
time is short compared to the parallel-shift time, but long 
compared to the serial-shift time (Cawley et al 2001).

The serial trails we see in the WFC are so short that they will have 
essentially no impact on aperture photometry, even for the smallest 
conceivable apertures.  It is therefore not surprising that it has not 
yet been observed in photometry (Riess \& Mack ISR 2004-06).

Since it is important to verify that the short trails we see do in 
fact contain all of the flux, meaning that no photometric correction
will be necessary, we examined the photometric residuals against $x$ 
as we had measured them against $y$.  We found no significant signature 
of serial CTE losses in the photometry.  Clearly almost all of the 
serial-CTE trails are extremely local.

The local nature of these trails does not, however, mean that it is not
important to correct them.  Since the trails have different amplitudes
as a function of pixel value, this will have a disparate impact on stellar
astrometry:  the faint stars will appear to be shifted by $\sim$0.01 pixel
away from the readout amplifier, while the brighter stars will appear 
to be shifted by only $\sim$0.004 pixel.  (We infer this by simply reading 
off the first-pixel fractions from Figure~\ref{fig25}).  For a given 
star, this may be within most margins of error, but projects that 
involve analyzing the motions of many stars to determine the average 
motion for one population relative to another can, in principle, get 
precisions below 0.001 pixel.  Such studies will be very susceptible 
to X-CTE biases.  The prescription we suggest for parallel CTE in the 
preceding analysis should also be amenable to CTE correction in the 
serial direction.   Indeed, since the trails are much more local, 
it should be much simpler to converge on the model parameters.

\clearpage


\section{SUMMARY AND NEXT STEPS}
\label{s.SUMMARY}

It has not been the goal of this document to create the definitive
black-box prescription for fixing CTE losses in ACS/WFC images.  
Rather, our aim has been simply to demonstrate a proof-of-concept.  
There are clearly many improvements that can be made, and we will 
anticipate some of them below.

\subsection{Accomplishments of the model}
\label{ss.SUMMARY_DONE}
We have shown that it is indeed possible to simulate the readout 
mechanism for a CCD detector by means of a simple empirical model.  
This simulation can then be used to compute for us the probable 
initial distribution of charge amongst the pixels for a given 
read-out image.  

Our model addresses the two long-standing concerns about imperfect
CTE:\ \  the removal of the trails downstream of bright sources 
and the restoration of flux to the pixels where it originated.  The 
algorithm was tailored to remove the trails from warm pixels in 
dark exposures, but we have shown (\S\ref{ss.MODEL_BKGD}) that 
it works equally well for backgrounds up to $\sim$35 $e^{-}$, with 
no reason to think it would not work for higher backgrounds, 
where the CTE impact is lower.  We have shown that when it removes 
the trails from stars (\S\ref{ss.STARS_BYEYE}), it puts all the 
flux back into the right place so that the star is measured to 
have the right brightness and the right position, as well 
(\S\ref{ss.STARS_SHvDEEP} and \S\ref{ss.OFFICIAL_COMP}).  
We even showed in \S\ref{ss.STARS_SHAPE} that restored stars 
have the right shape, as well.

One of the shortcomings of any deconvolution procedure is its 
sensitivity to noise.  Read-out noise, in particular, arises after
the charge-transfer errors have been imprinted on the pixels, so
we want to be certain that our algorithm does not over-correct
artifacts that are unrelated to CTE.  Our simulations show that
by applying our algorithm blindly, we could increase the effect
of read-out noise by almost 40\%.  We have devised a conservative
mechanism whereby our procedure only operates on the structure in
the image that is clearly consistent with not being readnoise.
This procedure reduces the impact of readnoise amplification to 
less than 3\%.

We have used the same procedure to examine CTE in the serial
direction.  Traps in the serial register are extremely short-lived.
They affect at most 1\% of the charge in a pixel and release the
charge within one transfer.  This will have a negligible effect on 
photometry, but will have a small but non-negligible effect on 
astrometry.  It should be possible to correct for serial CTE 
losses in a similar manner to our parallel correction.

The model we have developed clearly has the flexibility to address 
the observable aspects of CTE, even though it is silent on the
unobservable aspects, such as what happens at each specific stage 
of the three-phase pixel shift or exactly how many trap species with
different release profiles there are.  These unobservable details 
may have a lot to do with what actually leads to charge-transfer 
errors, but it is possible to address CTE issues without answering 
these questions directly.  Perhaps the clarity of our model will help 
experts develop a more detailed theoretical understanding of the 
phenomenon, and this improved understanding will surely yield even 
better corrections.

The model has been parametrized to allow a straightforward hand-fitting 
to the particulars of any detector.  This approach can and should be 
attempted on other HST instruments, such as the HRC, WFPC2, STIS, and 
perhaps even WFC3 once the CTE errors rise to a measurable level.
There may also be other non-HST detectors for which this model might 
prove useful.
 
\subsection{Future Plans}
\label{ss.SUMMARY_TODO}

There remain many avenues yet to pursue.  This paper has focused
exclusively on post-SM4 observations, where the CTE losses are as 
large as they have ever been.  It should be straightforward to extend 
this analysis to earlier epochs by simply scaling down the amplitude 
of our parameters.  While the WFC's electronics board was replaced, the
readout timing is unchanged and we expect our CTE model to work
with essentially the same parameters for pre-SM4 data.  This will
need to be demonstrated, however.  It will also be important to 
examine WP trails to study how CTE may or may not be impacted
by anneals and temperature changes.  Laboratory experiments 
(Janesick 2001) have shown that the trap-release time is faster 
for higher temperatures, so perhaps the trails will be even easier 
to model before the temperature was changed from $-77^{\circ}$ C 
to $-81^{\circ}$ C in 2006.

Our model has treated each pixel as being identical to all the
others in terms of having the same distribution of electron-capturing
traps.  Since the traps are presumed to result from individual 
radiaton-damage events, in actuality the pixel array will have a finite 
number of traps, each of which affects a specific level of charge 
packet.  On average, every other pixel will be impacted by a trap at 
one depth or another, such that we might expect 1000 $\pm \sqrt{1000}$ 
traps in each 2048-pixel column.  A variation of 3\% would not seem 
important.  However, if the trap-creating events generate multiple traps 
at once, as is expected for energetic neutron or ion impacts, 
then we might expect more variation than this from column to column, which 
might make our corrections less effective.  We attempted to look into 
this by examining the corrections of CR trails, but did not see a 
significant and consistent column-by-column effect.  Perhaps the best 
way to evaluate this would be to examine the efficacy of the photometric 
correction for stars of known brightness as a function of column 
number.  With the many sets of short+deep exposures in the archive 
of fields with good numbers of stars, it should be possible to rack 
up hundreds of star observations in each column.

The model we constructed is based on a representation of the density 
of traps that affect electron packets with various amounts of charge. 
It was difficult to constrain the trap density at the charge extremes.
There were very few warm pixels with more than 10,000 DN$_2$ in the
$\sim$1000 s integration.  As such, we had relatively few constraints 
on the density of traps that affect the electron packets at the bright 
end.  This node in the $\Phi_Q$ array might be better constrained by 
studying stars in short+deep combinations of images.  

At the low-intensity extreme, it was hard to identify warm pixels with 
fewer than 10 DN$_2$ at the top of the register, since this is so close 
to the readnoise level and since these low WPs are so smeared out by 
CTE losses.  These issues could be further investigated using warm 
pixels closer to the register, where the expected number of traps 
encountered per electron might be much less than one.  It would be 
good to understand the low end of the density profile, even though 
there may still be nothing that can done to restore the image.  It 
would also be good to take some shorter dark exposures, so that we can 
use our knowledge of the true intensity of the WPs from the deep darks 
to study how CTE impacts the smaller electron packets.  Also, with 
shorter darks, the crowding of WPs at the low-DN$_2$ level would be 
significantly lessened.  The ACS team is looking into taking darks 
with a variety of exposure times to explore these issues.

The model itself is designed to treat CTE losses as a perturbation.
We adopted the simplification that it is possible to treat the
transfer of flux from a pixel 2000 pixels from the register as having
simply 2000$\times$ CTE losses of a single transfer.  Essentially, we
are assuming that $ ( CTE )^N \approx 1 - N \times ( 1 - CTE )$.
This assumption should 
be revisited.  While it would be prohibitive to treat individually 
each and every one of the 2048$\times$1024 pixel-to-pixel shifts that 
take place in each column\footnote{
    The typical pixel is shifted down 1024 rows, and there are
    2048 pixels in each column.}, it should be possible to find a happy 
medium where we can still safely treat imperfect CTE as a perturbation 
within each modeled transfer.  One solution to this would be to run 
the algorithm three times in succession, using only one-third the 
current-model scaling.  This would ensure that CTE losses are indeed 
treated as a small perturbation within each iteration.  We have begun 
experimenting with this, but are awaiting shorter dark exposures to provide 
more useful tests.  One early result worth mentioning is that in this 
iterative scheme, where losses are kept at the perturbation level, the 
parameters of our model will naturally change somewhat, and we find that 
the trail profiles for the low-intensity WPs become more like those for the 
high-intensity trails, such that all WPs have the same trail profile.  
This result makes sense in light of the explanation in (Hardy et al.\ 1998) 
about how the traps that impact low- and high-intensity packets 
differ only in terms of their location within the lattice relative 
to the pixel boundaries.  As such, their release profiles should 
sensibly be the same.  This will be pursued in future work (Paper II), 
once the shorter darks are in hand. 

Finally, the possibility of correcting for imperfect CTE at the source 
could have profound implications for how best to operate the ACS pipeline.  
Should a CTE algorithm such as this one be executed on images before 
they are pipeline-processed, or is this best done after-market, and 
only by those who really need it?  Might the optimal approach depend 
on the science goals?  To highlight these complications, we note that 
our entire analysis has been premised on the fact that the dark images 
that we downloaded do {\it not\,} really reflect just the raw dark current 
in each pixel.  Rather, the true dark-current image has clearly been 
smeared out by the CTE-afflicted readout process.  Science images will 
suffer differing amounts of warm-pixel smearing depending on their 
background and exposure time; as such there is no single dark image 
that can be scaled up to estimate the impact of dark current and warm 
pixels on each pixel for any integration.  Addressing this properly will 
require much thought, but there probably is not a single ``right'' way 
to deal with this.  Bristow (2004) fleshes out some of the issues 
involved in considering such a CTE-pre-processor for STIS.

An additional consideration that arises when dealing with this correction 
as a part of the pipeline is what should be don3 with the error image that 
is provided along with the calibrated science image.  The CTE correction 
cannot be done perfectly, and the corrected image may have some systematic 
errors related to the imperfection of the algorithm and some random 
errors related to the fact that any deconvolution-type algorithm tends 
to amplify readnoise.  It will be important to run astrometric and
photometric tests on standard fields to come up with an effective way 
to represent these errors.
 
We have provided the details of our algorithm in this report, but have 
not yet made public the actual code.  This is because the algorithm is
currently being evaluated by the Institute to determine whether and
how best to implement it in the various data products.  Some version
of it should be available shortly.

\acknowledgements

We acknowledge Linda Smith, ACS Team Lead, and David Golimowski for
helpful discussions and encouragement.  We are also grateful for many
invaluable discussions about CTE with Paul Goudfrooj, Andy Fruchter, 
Jennifer Mack, Adam Riess, Stefano Casertano, Marco Chiaberge, and 
many others.   We are grateful to the anonymous referee for pointing 
us to several articles that discuss results of laboratory testing.  
Such articles often do not show up on astronomical search engines, 
and were glad to be made aware of them.  Finally, J.A.\ acknowledges 
Oleg Gnedin (PI of GO-10824 and GO-11589) for his patience in waiting 
for the hyper-velocity-star proper-motion measurements that this 
correction should make possible.

\references

{ 
\parindent -0.10in
\narrower

Anderson, J. \& King, I. R.  2000 PASP 112 1360

Anderson, J. \& King, I.\ R. 2006,  ACS/ISR 2006-01,  
    {\it PSFs, Photometry, and Astrometry for the ACS/WFC}

Anderson, J. 2007, ACS/ISR 2007-08,
    {\it Variation of the Distortion Solution of the WFC}

Anderson, J., Piotto, G., King, I.\ R., Bedin, L.\ R., 
    Guhathatkurta, P. 2009, ApJ, 697L, 58, 
    {\it Mixed Populations in Globular Clusters. Et Tu, 47 Tuc?}

Biretta, J. \& Mutchler, M.  1997 ISR WFPC2 97-05
    {\it Charge Traping and CTE Residual Images in the WFPC2 CCDs}

Biretta, J. \& Kozhurina-Platais, V.\  2005 WFPC2 ISR/05-01 
    {\it Hot Pixels as a Probe of WFPC2 CTE Effects}

Bristow, P. \& Alexov, A. 2002, CE-STIC-2002-001, 
    {\it Modelling Charge-Coupled Device Readout:  Simulation
     Overview and Early Results}

Bristow, P. 2003a CE-STIS-2003-001,  
    {\it Application of Model-Derived Charge Transfer 
         Inefficiency Corrections to STIS Photometric CCD Data}

Bristow, P. 2003b CE-STIS-2003-002, 
   {\it Application of Model-Derived Charge Transfer Inefficiency 
    Corrections to STIS Spectroscopic Data}

Bristow, P.  2004 CE-STIS 2004-002, 
   {\it The Impact of the CTI-correction Pre-Processor upon
        Standard STIS CCD Pipeline Dark and Bias Reference Files}

Cawley, L., Goudfrooij, P., Whitmore, B., Stiavelli, M., et al. 
     2001- WFC3 ISR 2001-05, {\it HST CCD Performance in the 
     Second Decade:  Charge Transfer Efficiency}

Chiaberge, M., Lim, P.-L., Kozhurina-Platais, V., 
    Sirianni, M. \& Mack, J.  2009, ACS/ISR 2009-01,
    {\it Updated CTE Photometric Correction for WFC and HRC}

Dolphin, A.\ E. 2000, PASP, 112, 1397

Dolphin, A.\ E. 2009, PASP, 121, 655

Golimowski, D. et al.\ 2010, ISR in prep.

Goudfrooij, P., Bohlin, R. C., Ma\'iz-Appellan\'iz, J. \& Kimble, R. A.
    2006, PASP, 848, 1453

Grogin, N. et al.\ 2010, ISR in prep.

Hardy, T., Murowinski, R., \& Deen, M. J. 1998
    IEEE Transaction on Nuclear Science Vol 45 N2 1998 p154

Janesick, J., Soli, G., Elliott, T., \& Collins, S.  1991 
    SPIE Vol 1447 Charge-Coupled Devices and Solid State 
    Optical Sensors II 

Janesick, J. R. Scientific Charge-Coupled Devices 2001
     SPIE Press monograph, Bellington, Washington.

Kozhurina-Platais, V., Goudfrooij, P., \& Puzia, T. H. 
    2007 ACS/ISR 2007.04,
    {\it ACS/WFC: Differential CTE corrections for Photometry and
         Astrometry from non-drizzled images}

Massey, R., Stoughton, C., Leauthaud, A., Rhodes, J., Koekemoer, A., 
    Ellis, R., Shaghoulian, E.  2010, MNRAS, 401, 371	

Maybhate, A. et al.\ 2010  
    ACS Instrument Handbook, Version 9.0 (Baltimore: STScI) 

McLaughlin, D.\ E.,  Anderson, J.,  Meylan, G., Gebhardt, K.,  Pryor, C,  
    Minniti, D., \& Phinney, S.  2006, ApJS, 166, 249
    {\it Hubble Space Telescope Proper Motions and Stellar Dynamics 
         in the Core of the Globular Cluster 47 Tucanae}
   
Mutchler, M. \& Sirianni, M. 2005, ACS/ISR 2005-03,
    {\it Internal Monitoring of ACS Charge Transfer Efficiency}

Philbrick, R. H.,  Ball Aerospace \& Technologies Corp. 2001,
   {\it Modelling the Impact of Pre-flushing on CTE
        in Proton Irradiated CCD based Detectors}

Riess, A. 2000, WFPC2/ISR 2000-04, 
   {\it How CTE Affects Extended Sources}

Riess, A. \& Mack, J.  2004, ACS/ISR 2004-06,
    {\it Time Dependence of ACS WFC CTE Corrections for Photometry 
     and Future Predictions}

Sirianni et al 2007, IEEE Radiation Effects Data Workshop
    {\it Radiation Damage in Hubble Space Telescope Detectors}

Waczynski et al 2001
     IEEE Transaction on Nuclear Science Vol 48 N 6 2001 pag 1807

}

\bigskip
\bigskip

\clearpage 
\end{document}